\documentclass[12pt]{article}

\usepackage{psfig}
\usepackage[dvips]{graphicx}
\usepackage{tabularx}
\usepackage{latexsym}
\usepackage{amsmath,amssymb,exscale}
\usepackage{array,multicol}
\numberwithin{equation}{section}
\usepackage{amsmath}
\usepackage{amssymb}

\newcommand{\nc}{\newcommand}
\newcommand{\rnc}{\renewcommand}

\rnc{\baselinestretch}{1.24}    
\rnc{\arraystretch}{1.24}   

\nc{\be}{\begin{equation}}

\nc{\ee}{\end{equation}}

\nc{\bea}{\begin{eqnarray}}

\nc{\eea}{\end{eqnarray}}

\nc{\ben}{\begin{eqnarray*}}

\nc{\een}{\end{eqnarray*}}

\nc{\xx}{\nonumber\\}

\nc{\ct}{\cite}

\nc{\la}{\label}

\nc{\eq}[1]{(\ref{#1})}

\nc{\newcaption}[1]{\centerline{\parbox{6in}{\caption{#1}}}}

\nc{\fig}[3]{
\begin{figure}
\centerline{\epsfxsize=#1\epsfbox{#2.eps}}
\newcaption{#3. \label{#2}}
\end{figure}
}

\def\IR{{\hbox{{\rm I}\kern-.2em\hbox{\rm R}}}}
\def\IB{{\hbox{{\rm I}\kern-.2em\hbox{\rm B}}}}
\def\IN{{\hbox{{\rm I}\kern-.2em\hbox{\rm N}}}}
\def\IC{\,\,{\hbox{{\rm I}\kern-.59em\hbox{\bf C}}}}
\def\IZ{{\hbox{{\rm Z}\kern-.4em\hbox{\rm Z}}}}
\def\IP{{\hbox{{\rm I}\kern-.2em\hbox{\rm P}}}}
\def\IH{{\hbox{{\rm I}\kern-.4em\hbox{\rm H}}}}
\def\ID{{\hbox{{\rm I}\kern-.2em\hbox{\rm D}}}}

\def\det{{\rm det}}

\title{
\begin{flushright}
\normalsize{CERN-PH-TH/2005-081\\IP/BBSR/2005-3 \\ 
\texttt{hep-th/0505260}}\\
\end{flushright}
\vspace*{6mm}
{\Large\textbf{Moduli stabilization with open and closed string 
fluxes}}
\author{\bf\large{
Ignatios Antoniadis$^{1}$\footnote{On leave from CPHT (UMR du CNRS 
7644), 
Ecole Polytechnique, F-91128 Palaiseau 
Cedex.}~\footnote{Ignatios.Antoniadis@cern.ch}~,
Alok Kumar$^{2}$\footnote{kumar@iopb.res.in}~, 
Tristan Maillard$^{1,3}$\footnote{Tristan.Maillard@cern.ch}}\\  
\\[-3mm]
\emph{\normalsize $^1$Department of Physics, CERN - Theory Division, 
CH-1211 Geneva 23, Switzerland }\\
\emph{\normalsize $^2$Institute of Physics, Bhubaneswar 751 005, 
India}\\
\emph{\normalsize $^3$Institut f\"ur Theoretische Physik, ETH 
H\"onggerberg, CH-8093\, Z\"urich, Switzerland}
}}

\date{}
\begin{document}
\maketitle
\thispagestyle{empty}


\begin{abstract}
We study the stabilization of all closed string moduli in the 
$T^6/\mathbb{Z}_2$ orientifold, using constant internal 
magnetic fields and 3-form fluxes that preserve ${\cal N}=1$ 
supersymmetry in four dimensions. We first analyze the 
stabilization of K\"ahler class and complex structure moduli 
by turning on magnetic fluxes on different sets of $D9$ branes 
that wrap the internal space $T^6/\mathbb{Z}_2$. We present
explicit consistent string constructions, satisfying in particular
tadpole cancellation, where the radii can take arbitrarily large 
values by tuning the winding numbers appropriately.
We then show that the dilaton-axion modulus can also be fixed by
turning on closed string constant 3-form fluxes, consistently with
the supersymmetry preserved by the magnetic fields, providing at 
the same time perturbative values for the string coupling. 
Finally, several models are presented combining 
open string magnetic fields that fix part of K\"ahler class and 
complex 
structure moduli, with closed string 3-form fluxes that stabilize the 
remaining ones together with the dilaton.
\end{abstract}

\date


\newpage

\section{Introduction}
String theory is known to possess a large number of vacua which 
contain the basic structure of grand unified theories, and in 
particular 
of the Standard Model. 
However, one of the major stumbling blocks in making 
further progress along these lines has been the lack of a guiding
principle for choosing the true ground state of the theory, thus 
implying the loss of predictivity.
In particular, string vacua depend in general on continuous 
parameters, characterizing for instance the size and shape of the
compactification manifold,
that correspond to vacuum expectation values (VEVs) of the 
so-called moduli fields. These are perturbative flat directions of 
the 
scalar potential, at least as long as supersymmetry remains unbroken.
It is therefore of great interest that during the last few years
there has been a considerable success in fixing the string ground 
states, by invoking principles similar to the spontaneous 
symmetry breaking mechanism, now in the context of string theory.
In particular, it has been realized that closed, as well as open,
string background fluxes can be turned on,
fixing the VEVs of the moduli fields and therefore
providing the possibility for choosing a ground state as a local
isolated minimum of the scalar potential of the theory.
This line of approach allows string theory to play directly a role 
in particle unification, predicting the strength of interactions and
the mass spectrum. In particular, the string coupling becomes a
calculable dynamical parameter that fixes the value of the fine
structure constant and determines the Newtonian coupling in 
terms of the string length.

On one hand, moduli stabilization using closed string 
3-form fluxes has been discussed in a great detail in the 
literature \cite{Frey:2002hf, Review}. 
${\cal N}=1$ space-time supersymmetry and
various consistency requirements imply that the 
3-form fluxes must satisfy the following conditions formulated
on the complexified flux defined as $G = F - \phi H $, 
where $F$ and $H$ are the R-R (Ramond) and
NS-NS (Neveu-Schwarz) 3-forms, respectively, 
and $\phi$ is the axion-dilaton modulus: 
(1) The only non-vanishing components of $G$ are of the type $(2,1)$, 
pointing along two holomorphic and one anti-holomorphic directions,
implying that its $(1,2)$, $(3,0)$ and $(0,3)$ components
are zero and (2) $G$ is primitive, requiring $J \wedge G = 0$
with $J$ being the K\"ahler form. 
This approach has been applied to orientifolds of both 
toroidal models as well as of Calabi-Yau compactifications. 
However, a drawback of the method is that the K\"ahler class 
moduli remain undetermined due to the absence of an harmonic $(1,0)$ 
form on Calabi-Yau spaces, implying that the constraint 
$J \wedge G = 0$ is trivially satisfied. 
In the toroidal orientifold case, it turns out that one is
able to stabilize the K\"ahler class moduli only partially,
but in particular the overall volume remains always unfixed.

On the other hand, in \cite{AM} two of the present authors have 
shown that both complex structure and K\"ahler class moduli can 
be stabilized in the type I string theory compactified down to four 
dimensions.\footnote{For partial K\"ahler moduli stabilization, see 
also
\cite{Blumenhagen:2003vr, Cascales:2003zp}.} 
This is achieved by turning on magnetic fluxes which 
couple to various $D9$ branes, that wrap on $T^6$, through a 
boundary term in the open string world-sheet action. The latter 
modifies 
the open string Hamiltonian and its spectrum, and puts 
constraints on the closed string background fields due to their 
couplings to the open string action. 
More precisely, supersymmetry conditions in the presence of 
branes with magnetic fluxes, together with conditions which define 
a meaningful world-volume theory, put restrictions on the values of 
the moduli and fix them to specific constant values.
This also  breaks the original ${\cal N}=4$ supersymmetry of the 
compactified type I theory to an ${\cal N}=1$ supersymmetric gauge 
theory with a number of chiral multiplets. A detailed analysis of the 
final spectrum, as well as other related issues have been discussed 
in \cite{Bianchi:2005yz}. 

In the simplest case, the above model has only $O9$ orientifold 
planes and several stacks of magnetized $D9$ branes. The main
ingredients for moduli stabilization are then: (1) the introduction
of ``oblique" magnetic fields, needed to fix the off-diagonal
components of the metric, that correspond to mutually 
non-commuting matrices similar to non-abelian orbifolds;
(2) the property that magnetized $D9$ branes lead to negative
5-brane tensions; and (3) the non-linear part of Dirac-Born-Infeld 
(DBI) action which is needed to fix the overall volume.
Actually, the first two ingredients are also necessary for
satisfying the 5-brane tadpole cancellation without adding
$D5$ branes or $O5$ planes, while the last two properties
are only valid in four-dimensional compactifications (and not
in higher dimensions). 

In this paper we construct ${\cal N}=1$ supersymmetric models 
with stabilized moduli in $T^6/\mathbb{Z}_2$ orientifold 
compactifications of type IIB theory, following the earlier 
work in \cite{AM}. In the simplest case, our models have 
only $O3$ orientifold planes and several stacks of magnetized 
$D9$ branes that behave as $D3$ branes. The induced 
7-brane tadpoles cancel without the addition of extra $D7$ 
branes or $O7$ planes. We write down the relevant 
supersymmetry requirements, and demand that  the 
world-volume theory should be well defined.
We then analyze these conditions for several situations,
to examine what magnetic 
fluxes can be turned on along the $D9$ branes consistently.
One may think that the results of this work can be obtained
simply by a T-duality from the toroidal case with $O9$ planes
analyzed in \cite{AM}. This is indeed true only for
invertible magnetic field matrices. On the contrary, if the magnetic
flux has a zero eigenvalue, in the T-dual theory it becomes
infinite and the analysis does not go through. Thus, the study
of moduli stabilization in the $T^6/\mathbb{Z}_2$ orientifold case is
non-trivial and cannot be obtained by a T-duality from the
toroidal analysis of ref.~\cite{AM}.

Actually, we are interested to find an explicit solution, where
the toroidal geometry of $T^6$ is fixed to a factorized form,
$T^6 \equiv {(T^2)}^3$. Thus, one needs in particular to set 
the off-diagonal components of the complex structure to zero, implying
the presence of magnetic fluxes with non-zero 
off-diagonal components, mixing the $T^2$'s as in \cite{AM}.
However, unlike that case, the consistency conditions
now imply that one has to simultaneously turn on certain 
diagonal (in $T^2$'s) fluxes as well.
Concerning the branes with purely diagonal form of magnetic 
flux along the three $T^2$'s, we find that it
is allowed to have  a zero flux along one $T^2$
and two negative fluxes along the remaining two tori.
In such situtations, it is also possible to turn on 
off-diagonal components of magnetic fluxes in the directions 
orthogonal to the $T^2$ with zero flux.
In fact, we make use of such purely diagonal fluxes, as well
as the ones with off-diagonal components, since they all
provide conditions on moduli without contributing
to the 3-brane tadpoles.

The restrictions we find in this paper on the possible allowed 
fluxes, turn out to be more restrictive than in \cite{AM}.
Nevertheless, we have been able to use them for the purpose of  
stabilizing all the complex structure and K\"ahler class moduli.
In fact, additional restrictions on the string construction emerge
from the requirement of 7-brane and 3-brane tadpole cancellations. 
As mentioned above, the contribution of 7-brane 
tadpoles depends on the $D9$ brane winding around the 
corresponding transverse 2-cycles. It turns out that the 7-brane 
tadpole 
contribution within a stack of branes can take positive or negative 
values along the various 2-cycles. 
On the other hand, the 3-brane tadpole contribution within a stack 
of branes is not affected by the windings, and is restricted to be 
positive. We then keep them to their minimum positive value in 
order to have the possibility of introducing closed string 3-form 
fluxes as well, so as to finally fix the only remaining closed string
moduli field, namely the axion-dilaton modulus. 
Indeed, we are able to find consistent models within this framework
where the string coupling is fixed to perturbative values.
By tuning appropriately the magnetic fluxes,
we also find an infinite but discrete series of solutions with
stabilized moduli, where some radii can take arbitrarily large
values and the dilaton can be fixed at arbitrarily weak coupling. 
We finally present models where part of the K\"ahler and 
complex structure moduli are stabilized using the closed string 
3-form flux and the other part by open string magnetic fluxes.
In these cases, we are able to obtain even smaller values for 
the string coupling.

In this work, we do not address the issue of open string moduli
stabilization. In particular, we study only vacua where gauge
symmetries are unbroken. If one allows the possibility of gauge
symmetry breaking, other vacua should exist where K\"ahler 
moduli mix with open string D-term flat directions and thus only
one linear combination is fixed by the presence of the corresponding 
magnetic
field \cite{Antoniadis:1997mm}. In principle, the remaining 
directions 
can be also fixed by adding more magnetic fields but such 
an analysis goes beyond the scope of the present paper.

The rest of the paper is organized as follows. In Section 
\ref{sec:SUSY}, we 
write down the consistency conditions for magnetic fluxes on 
$D9$ branes in $T^6/\mathbb{Z}_2$ orientifold models, leaving
unbroken ${\cal N}=1$ supersymmetry in four dimensions. 
Supersymmetry conditions are analyzed in subsection \ref{sec:SUSY2}, 
while
tadpole cancellation and positivity requirements are discussed
in subsection \ref{sec:tadpoles}. We also describe the general 
mechanism for
moduli stabilization in subsections \ref{sec:moduli_stab} and 
\ref{sec:rrmod}. The notations are
the same as in ref.~\cite{AM} but for self-consistency, in Appendix 
\ref{notations},
we present briefly the torus $T^6$ parametrization.
In Section \ref{sec:3ff}, we review the supersymmetry and consistency 
conditions of closed string 3-form fluxes and discuss
the effects of turning on a non-trivial NS-NS $B$-field background.
In Section \ref{b&f}, we give in advance the various brane stacks and
choices for the magnetic fluxes that will be used in the examples of
string constructions of the following sections. In Section 
\ref{explicit}, we
present an explicit model in detail (called model-A), using
twelve magnetized $D9$ branes, contributing
the lowest possible value to the 3-brane tadpole, $q_{3,R}=6$.
We show that our choice of magnetic fields satisfies all consistency
requirements, leading to a ${\cal N}=1$ supersymmetric vacuum where
all complex structure and K\"ahler class moduli are fixed and the
metric becomes diagonal in the internal coordinates. In the
subsection \ref{sec:A-3-form}, we also show that the dilaton-axion 
modulus is
also fixed by turning on appropriate 3-form fluxes at weak values
of the string coupling. Therefore, all closed string moduli get
fixed. Finally, in the last subsection \ref{subsec:nongeom}, we 
present a possible alternative based on a minimal number of nine
magnetized $D9$ branes, leading to an infinite (but 
discrete) family of solutions with the same
values of all geometric moduli, but with different spectrum and 
couplings.\footnote{This model however satisfies weaker constraints and
further work is needed to establish its consistency.} 
In Section \ref{sec:led}, we show how the above solution can be
`rescaled' to generate large values for (some of) the internal 
radii \cite{ld}. In Section \ref{sec:Model-B}, we repeat the analysis 
for another example
(called model-B), which uses 15 magnetized $D9$ branes
contributing 12 units of 3-brane charge, $q_{3,R}=12$. 
Many technical details of the model,
such as the choice of fluxes and windings, as well as the tadpole 
cancellation conditions are given in Appendix B. In Section 
\ref{sec:CD}, we
present a different model (called model-C), where closed string 
3-form fluxes are
also used to stabilize part of the geometric moduli, besides the
dilaton. In this way, the number of magnetized branes and their
contribution to the 3-brane tadpole is lower than before. Finally,
Section \ref{sec:conc} contains a brief summary of our results.

\section{Magnetic fluxes and supersymmetric vacua}\label{sec:SUSY}

\subsection{Condition for $\mathcal{N}=1$ 
supersymmetry}\label{sec:SUSY2}

The presence of a constant internal magnetic field generically breaks 
supersymmetry by shifting the masses of the four dimensional bosons 
and fermions \cite{Bachas:1995ik}. However, for suitable choice of 
the fluxes and 
moduli, a four dimensional supersymmetry can be recovered 
\cite{Angelantonj:2000hi}. 
Written in the complex basis (\ref{complex_structure}) of  Appendix 
\ref{notations}  where the  
field 
strength $\mathcal{F}$ splits in purely (anti-) holomorphic  
($\mathcal{F}_{(0,2)}$), $\mathcal{F}_{(2,0)}$ and mixed 
$\mathcal{F}_{(1,1)}$ parts, the condition for  $\mathcal{N}=1$ 
supersymmetry in four dimensions can be written 
as~\cite{Marino:1999af}:
\bea
(iJ+\mathcal{F})^{3} & = &e^{i\theta} 
\sqrt{|g_6+\mathcal{F}|}\frac{V_6}{\sqrt{|g_6|}} 
\label{kc1}\\
\mathcal{F}_{(2,0)}  & = & 0\, ,
\label{csc1}
\eea
where $V_6$ is the volume form of $T^6$ and $g_6$ is its metric. 
Eq.~(\ref{kc1}) can be put in the form:
\be
\tan{\theta}\left(J\wedge J \wedge \mathcal{F} - 
\mathcal{F}\wedge\mathcal{F} \wedge \mathcal{F}\right) = 
J \wedge J \wedge J - J\wedge 
\mathcal{F} \wedge \mathcal{F}\, ,
\label{kc2}
\ee
where the wedge product $A^N$ is defined with an implicit 
normalization factor $1/N!$.
Note that only the $(1,1)$-part of $\mathcal{F}$ contributes in this 
formula.  Formally, (\ref{kc2}) can be also written as
\be
\rm{Im}\left(e^{-i\theta}\Phi\right)=0\, ,
\label{kc3}
\ee
with 
\be
\Phi = 
(iJ+\mathcal{F})\wedge(iJ+\mathcal{F})\wedge(iJ+\mathcal{F})\, .
\label{Phi}
\ee
The constant phase $\theta$ selects which supersymmetry the 
magnetized 
brane preserves. In the case of type I string theory, the 
supercharges 
preserved by the magnetic background field is consistent with the 
presence 
of the orientifold plane $O9$ for the choice of $\theta = 
-\frac{\pi}{2}$. 
Consider on the other hand the orientifold compactification 
$T^6/\mathbb{Z}_2$,
where the $\mathbb{Z}_2$ orientifold projection is given by  $\Omega 
R (-)^{F_L}$. 
This is a composition of the world-sheet parity $\Omega$ with the 
parity R on the 
torus $T^6$: $z_i \rightarrow -z_i$ and the spacetime left handed 
fermionic number 
$(-1)^{F_L}$. The orientifold projection has 64 
fixed points on $T^6$, giving rise to 64  
$O3$-planes. Each of them carries negative tension and charge and 
preserves 
a common 
supersymmetry with the magnetized $D9$ branes for the special choice 
of 
phase 
$\theta = 0$ \cite{Blumenhagen:2003vr}. The supersymmetry condition 
(\ref{kc3}) 
reduces then to the formula
\be
J \wedge J \wedge J - J\wedge \mathcal{F} \wedge \mathcal{F} = 0. 
\label{kcf}
\ee

\begin{figure}[tbp]
    \centering
    \includegraphics[height=6cm]{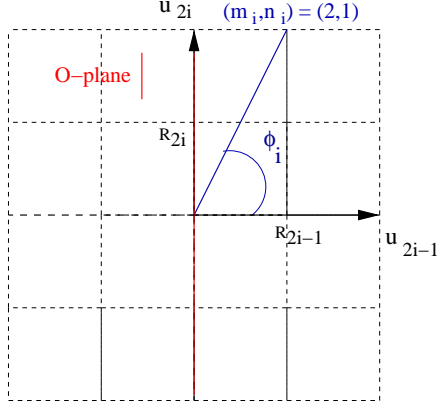}
    \caption{T-duality along the vertical axis $u_{2i}$. The 
$D$-brane 
dual to a magnetized brane forms an angle $\frac{\pi}{2}-\phi$ with 
respect to the vertical axis.  The 
orientifold plane, T-dual to an $O3$-plane,  sits along the vertical 
axis. }
    \label{fig:angle}
\end{figure}

The supersymmetry condition (\ref{kcf}) can also be 
understood in a type IIA T-dual representation in terms of 
the angles between different stacks of $D6$ branes. To illustrate 
this 
fact, let us 
consider a coordinate basis $u_{k}\,, \,\,\, k=1,\dots,6$, on 
the torus where 
the metric is the identity, $g_{kl}=\delta_{kl}$,
and the magnetic flux is block-diagonal $F = i\sigma_{2}\otimes 
(f_{1},f_{2},f_{3})$. We denote the 
radii of the coordinates as $ R_{k} $. The fluxes are then quantized 
as
\be
qf_{i}=\frac{m_{i}}{n_{i}R_{2i-1}R_{2i}}, \quad \quad i=1,2,3,
\ee
where $q$ is the quantum of the electric charge, $m_{i}$ are the 
first Chern numbers and $n_{i}$ are the 
winding numbers of the $D9$ brane around the cycles 
$[u_{2i}u_{2i-1}]$.
The boundary conditions of the open string coordinates in this 
magnetic background deform the pure 
Neumann 
conditions to 
\bea
\partial_{\sigma}u_{2i-1} -qf_{i}\partial_{\tau}u_{2i} &=& 0 \,\,\,\, 
{\rm{at}} \,\,\, \sigma = 0 \nonumber\\
qf_{i}\partial_{\tau}u_{2i-1} +\partial_{\sigma}u_{2i} &=& 0  
\,\,\,\, {\rm{at}} \,\,\,
\sigma = \pi \quad \quad i=1,2,3
\label{bc_magn}
\eea
where $\sigma $ and $\tau$ are the usual world-sheet coordinates. 
Upon three T-dualities along the directions $u_{2i}$,
\be
R_{2i} \rightarrow 1/R_{2i},
\ee
the boundary conditions are modified as
\bea
\partial_{\sigma}(u_{2i-1} -qf_{i}u_{2i} ) &=& 0 \,\,\,\,  {\rm{at}} 
\,\,\,
\sigma = 0 \nonumber\\
\partial_{\tau}(qf_{i}u_{2i-1} +u_{2i} ) &=& 0  \,\,\,\,  {\rm{at}} 
\,\,\,
\sigma = \pi \quad \quad i=1,2,3.
\label{bc_angle}\eea
The T-dualities also map the quantized $D9$ brane fluxes to the $D6$ 
brane angles 
\be
qf_{i}=\frac{m_{i}}{n_{i}R_{2i-1}R_{2i}} \rightarrow \tilde{f}_{i} = 
 \frac{m_{i}R_{2i}}{n_{i}R_{2i-1}}=\tan\phi_{i}. 
\ee
In fact, the first Chern numbers $m_i$ are mapped into the winding 
numbers of the 
$D6$ 
brane along the coordinates $u_{2i}$ while $n_{i}$ become  the 
winding numbers along the directions $u_{2i-1}$. Furthermore, as the 
three 
T-dualities map the $O3$ planes into 
$O6$ planes sitting along the $u_{2i}$ axis, the new boundary 
conditions (\ref{bc_angle}) correspond then to a $D6$ brane wrapped 
on a 
3-cycle defined by the angles with respect to the $u_{2i-1}$ axis 
given by 
$\tan\phi_{i} = \tilde{f}_{i}$ (Figure \ref{fig:angle}). In these new 
variables, 
the supersymmetry condition (\ref{kcf}) reads
\be
\sum_{i}\phi_{i} = \frac{3\pi}{2} \,\,\, {\rm mod} 2\pi.
\ee
The sum over the angles defined with  respect to the vertical axis 
where the $O6$ 
plane sits is then zero, as argued above.

\subsection{Moduli stabilization}\label{sec:moduli_stab}

From now on, we  will focus our attention to the orientifold 
compactification $T^6/\mathbb{Z}_2$ with $\theta_{O3} = 0$.
Following our analysis of eqs.~(\ref{csc1}) 
and (\ref{kc2}), we have seen that a single magnetized 
$D9$ brane stack preserves $\mathcal{N}=1$ supersymmetry in four 
dimensions 
for a restricted closed string moduli space. As we will see below, if 
we 
introduce several magnetic fluxes in the world-volume of different 
stacks of
$D9$ branes, it will be possible to fix completely all closed string 
moduli but 
the dilaton.

As in  \cite{AM},  eqs.~(\ref{csc1}) 
and (\ref{kcf}) can be interpreted as 
conditions which fix the moduli in terms of the magnetic fluxes.
More specifically, we consider $K$ stacks of $N_{a}$  $D9$ branes, 
with 
$a = 1, \cdots , K$. We introduce on each stack a 
background magnetic field with constant field strength $F^{a}$ on 
the corresponding
world-volume and endpoint charge $q_{a}$.  The magnetic fields are
separately quantized, following the Dirac condition 
\cite{Blumenhagen:2000vk}
\be
q_a F^a_{kl} =2\pi\cdot \frac{m^{a}_{kl}}{n^{a}_{kl}}\equiv 2\pi\cdot 
p^{a}_{kl} \,\,\, , \  p^{a}_{kl} \in \mathbb{Q} \,\,\, ,\,\,\, a = 1 
, 
\cdots , K\, .
\label{diraca}
\ee

Written in the complex coordinates (\ref{complex_structure}), the  
field strength is decomposed in a purely holomorphic and mixed part. 

The supersymmetry conditions for each stack ask then for a vanishing 
purely holomorphic field strength:
\be
F_{(2,0)} = 0 \rightarrow \tau^{T} p^{a}_{xx} \tau - 
\tau^{T}{p^{a}_{xy}} - p^{a}_{yx}\tau + 
p^{a}_{yy} = 0,\, 
\label{M20_condition}
\ee
where the matrices $(p^{a}_{xx})_{ij}$, $(p^{a}_{xy})_{ij}$ and 
$(p^{a}_{yy})_{ij}$
enter in the quantized field strength (\ref{diraca}) in the directions
$(x^i,x^j)$, $(x^i,y^j)$ and $(y^i,y^j)$, respectively, where $\tau$ 
is the complex structure matrix.\footnote{See parametrization in 
Appendix \ref{notations}.}
The second condition (\ref{kcf}) restricts the K\"ahler moduli to 
satisfy
\be
J \wedge J \wedge J - J\wedge 
\mathcal{F}^a \wedge \mathcal{F}^a = 0 \,\,\, ,\ a = 1 , \cdots , K\, 
. 
\label{kc1a}
\ee
We have used the fact that the phases $\theta_{a}$'s  of all stacks 
have to be the same in 
order for each stack to preserve the same 
supersymmetry: $\theta_{a} = 
0, \,\,\, \forall a=1,\dots,K$.
Furthermore, when the condition (\ref{M20_condition}) is fulfilled, 
the expression for the magnetized field strength $F$, denoted
 $\mathcal{F}$ in the complex basis (\ref{2fbasis}), reduces to the 
matrix:
\be
\mathcal{F}^a=\left(
\begin{array}{cc}
0 & Y^a\\
{Y^a}^\dagger & 0\\
\end{array}
\right)\quad ;\quad
Y^a=\frac{1}{2}(2\pi)^2 \alpha' {\rm{Im}} {\tau^{-1}}^T
\left( p_{yx}^a - \tau^{T} p^a_{xx}\right)\, .
\label{11part}
\ee
This splits in the real and imaginary parts:
\bea
{\rm{Re}} Y^a &=& \frac{(2\pi)^2\alpha'}{2}{\rm{Im}} 
{\tau^{-1}}^T\left( p^{a}_{yx} - 
{\rm{Re}}\tau^{T}p^{a}_{xx}\right)\, ,  \\
{\rm{Im}}Y^a &=& -\frac{(2\pi)^2 \alpha'}{2}p^a_{xx}\, .
\label{11part2}
\eea
Inspection of eqs.~(\ref{M20_condition}) and 
(\ref{kc1a}) 
shows that for each stack of magnetized $D9$ branes, we have up to 
three complex conditions 
for the moduli of the complex structure, depending on the directions 
in which the fluxes are switched on, whereas only one real
condition can be set on the K\"ahler moduli. Therefore, in order to 
fix the 
K\"ahler 
moduli, we must add more stacks of branes compared to the ones 
needed to fix the same number of complex structure moduli and at 
least nine in order to fix them all.\footnote{As mentioned in the 
introduction, the above counting of conditions
holds for vacua with unbroken gauge symmetries, without open string 
moduli
switched on.}

\subsection{Consistency conditions}\label{sec:tadpoles}

The presence of constant internal magnetic field strength induces 
lower 
dimensional charges and tensions. In a consistent compactification, 
these have to be cancelled by the contribution of lower dimensional 
objects (branes or orientifold planes) or other kinds of fluxes (such 
as  3-form fluxes). 
In the case of a $T^6/\mathbb{Z}_2$ compactification where the 
supersymmetry conditions
(\ref{csc1}) and (\ref{kcf}) 
are satisfied, the Dirac-Born-Infeld (DBI) and  Wess-Zumino (WZ) 
action of  magnetized $D9$ branes read:
\bea
V_{DBI}  & = & -T_{9} \sum_{a=1}^{K}\,N_{a} 
\int_{\mathcal{M}_{10}^{a}}\sqrt{|g+\mathcal{F}^{a}|} 
\nonumber \\
         & = & -T_{9}\,\sum_{a=1}^{K}\,N_{a} 
\int_{\mathcal{M}_{4}}\sqrt{|g_4|}\int_{\mathcal{M}^{a}_{6}} 
 {\rm Re}\, [ e^{-i\theta_{a}}(iJ+\mathcal{F}^{a})^{3}] \nonumber 
\label{DBI}\\
          &=&  
-T_{9}\sum_{a=1}^{K}\,N_{a}\int_{\mathcal{M}_{4}}\sqrt{|g_4|}\int_{\mathcal{M}^{a}_{6}}
\bigg\{\, \mathcal{F}^{a}\wedge \mathcal{F}^{a} \wedge 
\mathcal{F}^{a}- J\wedge J \wedge \mathcal{F}^{a}\bigg\}
\nonumber\\
V_{WZ} & = & \mu_{9}\sum_{a=1}^{K}\,N_{a}\int_{\mathcal{M}^a_{10}} C 
e^{\mathcal{F}^{a}} 
\nonumber \\
       & = & \mu_{9}\sum_{a=1}^{K}\,N_{a}\int_{\mathcal{M}^a_{10}}
\bigg\{C_{4}\wedge \mathcal{F}^{a}\wedge \mathcal{F}^{a}\wedge 
\mathcal{F}^{a}+C_{8}\wedge \mathcal{F}^{a}\bigg\}  
\label{WZ}
\eea
where $T_9$ and $\mu_9$ are the $D9$ brane tension and R-R charge, 
respectively, 
while the integral over the internal manifold $\mathcal{M}^{a}_6$ 
takes 
into account the winding numbers $n_{kl}^a$ of the different branes. 
The terms involving the R-R potentials $C_{10}$ and $C_6$ terms do 
not 
appear in the WZ action as they are projected out by the 
orientifold projection.  

Consider now the real
basis $\omega_r$ of $H^2(T^6)$, with $r = 1, \cdots , h_2$, in which  
the quantization condition (\ref{diraca}) for the magnetic fluxes 
reads:
\be
\frac{1}{2\pi} q_aF^a_r = \frac{m^a_r}{n^a_r} = p^a_r\, 
.\label{dirac3}
\ee
We now define the quantity
\be
\mathcal{K}_{rst} = \int_{T^6} \omega_r \wedge  \omega_s  
\wedge\omega_t 
\label{Ksign}
\ee
which is a sign, following the orientation choice given  in 
(\ref{orientation}). 
The 3-brane R-R charge, $q_{3, R}$, coming from the first term of  
the last line of 
(\ref{WZ}), reads 
\be
q_{3, R}=\sum_{a=1}^{K}\,\, N_a \sum_{r,s,t}\,\, 
\mathcal{K}_{rst}m^a_r 
m^a_s 
m^a_t \, .
\label{q3R}
\ee
Since we start with a $T^6/\mathbb{Z}_2$ orientifold with $O3$ planes 
carrying $-16$ units 
of R-R charge, the R-R tadpole cancellation condition 
implies
\be
q_{3, R}=16\, . 
\label{t3}
\ee

The second set of  conditions comes from the induced 7-brane R-R 
charges, emerging from the second term of eqs.~(\ref{WZ}). For each 
2-cycle $C^{(2)}_t$ of the torus $T^{6}$, there is a localized 7-brane
charge, given by $q_{7,R}^{t}$:
\be
q_{7,R}^{t}  = \sum_{a=1}^{K}\, \, N_a\,\sum_{r,s}
\,
\mathcal{K}_{rst}\, n^a_r n^a_s 
m^a_t =: \sum_{a=1}^{K}\, N_a\, q^{a}_{t}\, .
\label{q7R}
\ee
In the $T^{6}/\mathbb{Z}_{2}$ compactification,   7-dimensional 
orientifold planes are absent and  the total 7-brane 
tadpole contribution must thus vanish
for any 2-cycle $t$:
\be
q_{7,R}^{t}  =0 \,\,\, , \,\,\, \forall t= 1,\cdots, h_2\, .
\label{t7}
\ee
As a result, we will impose the R-R tadpole cancellation conditions 
(\ref{t3}) and (\ref{t7}): 
$q_{3, R}=16$ and $q_{7,R}^{t}  =0$, together with the supersymmetry
constraints (\ref{csc1}) or equivalently 
(\ref{M20_condition}), and (\ref{kc1a}).

Furthermore, even if magnetized antibranes may preserve the same 
supersymmetry
as the orientifold $T^6/\mathbb{Z}_2$, satisfying a different 
condition than
(\ref{kc1a}) \cite{Marchesano:2004xz}, here we will consider 
only a setup without antibranes. In the 
T-dual picture of $D6$ branes at angles presented in the previous 
section, the $O6$ plane is located along the axis $u_{2i}$. Then, 
from Figure~\ref{fig:antibrane}, the image 
of a brane with quantum numbers $(m_{i},n_{i})$, $i=1,2,3$, under the 
orientifold projection is a brane with quantum numbers 
$(m_{i},-n_{i})$. 
Moreover, an antibrane $\bar{D}6$ is obtained by 
a rotation by an angle 
$\pi$ from the corresponding $D6$ brane in an odd number 
of cycles $[u_{2i-1}u_{2i}]$, corresponding to 
a brane with winding numbers $(-m_{i},-n_{i})$. Therefore, the 
absence of antibranes is expressed as a condition on the winding 
numbers along the $u_{2i}$ axis, or equivalently on the first Chern 
numbers:

\begin{figure}[tbp]
    \centering
    \includegraphics[height=6cm]{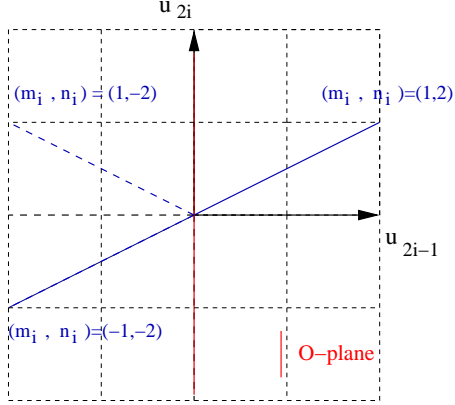}
    \caption{Example of a brane at angle with its orientifold image 
and antibrane. The $O$-plane is situated on the vertical axis. }
    \label{fig:antibrane}
\end{figure}

\be
K_{rst}m^{a}_{r}m^{a}_{s}m^{a}_{t} > 0 \quad \quad \forall a=1,\dots 
K.
\label{no_anti}
\ee
The limiting case where one of the first Chern numbers $m_{i}$ 
vanishes,
along the coordinates $u_{2i}$  
corresponds to the situation where the brane is horizontal in one of 
the 
 2-cycles $[u_{2i-1}u_{2i}]$. Switching the sign of the winding 
number $n_{i}$ corresponds then to switch a brane into an antibrane. 
The condition for the absence of antibranes in this case then reads:
\be
K_{rst}m^{a}_{r}m^{a}_{s}n^{a}_{t}>0\quad \quad \forall a=1,\dots K\, 
.
\label{no_anti2}
\ee

Next, a condition of positivity for the real part of  
$\Phi_{a}$ defined in eq.~(\ref{Phi}) has to be satisfied for each 
$a$,
as it corresponds to the generalized world-volume element of 
each separate brane stack:
\be
{\rm Re}(e^{-i\theta_a}\Phi_a) > 0\, , \quad \forall\, a 
= 1,\cdots , K\, ,
\ee
with
\be
\Phi_a = 
(iJ+\mathcal{F}^a)\wedge(iJ+\mathcal{F}^a)\wedge(iJ+\mathcal{F}^a)\, .
\label{Phia}
\ee
For $\theta_a=0$, it reduces to the condition:
\be
\mathcal{F}^a\wedge\mathcal{F}^a \wedge\mathcal{F}^a-J\wedge J 
\wedge\mathcal{F}^a > 0\, .
\label{Jcond}
\ee
Let us consider two cases which will arise in the examples of the 
following
sections.
\begin{itemize}
\item
When only the  diagonal K\"ahler form elements $J_{i}=:J_{i\bar{i}}$ 
are non-zero and all off-diagonal fluxes vanish $F_{i\bar{j}}=0$, the 
positivity 
condition (\ref{Jcond}), together with the supersymmetry condition 
(\ref{kc1a}), reads:
\be
-J_{1}J_{2}J_{3}(H_{2}+H_{3})\left(1+H_{1}^{2}\right)>0,
\label{Jcond1}
\ee
where we use the notation $F_{i} = F_{i\bar{i}}$ and $H_{i} = 
F_{i}/J_{i}$. As all K\"ahler moduli $J_{i}$ are volumes, 
they are  positive and the above condition becomes 
\be
H_{2}+H_{3}<0\, .
\label{Jcond2}
\ee
\item
In the last case we will consider, there are also 
non-diagonal fluxes, like for example $F_{1\bar{2}}$, together with a 
diagonal one  $F_{3}$. 
Eq.~(\ref{Jcond}) then reads
\be
-F_{3}|F_{1\bar{2}}|^{2}\left( 
1+\frac{J_{1}J_{2}}{|F_{1\bar{2}}|^{2}} \right) > 0, \label{Jcond3}
\ee
implying that the diagonal component $F_{3}$ has to be negative. 
\end{itemize}

Finally, we compute the intersection number $I_{ab}$ between
the stacks $a$ and $b$, which gives the number of chiral fermions. 
As it has been shown in
\cite{Bianchi:2005yz}, $I_{ab}$ in the presence of a
general magnetic flux can be written as 
\bea
I_{ab}& = & -\frac{W_aW_b}{(2\pi)^3}\int_{T^6}c_3(L_a\otimes 
L_b)\nonumber \\
      & = & W_a W_b \sum_{rst} K_{rst}      
\frac{(m^a_rn^b_r-m^b_rn^a_r)(m^a_sn^b_s-m^b_sn^a_s)(m^a_tn^b_t-m^b_tn^a_t)}      
{n^a_rn^a_sn^a_tn^b_rn^b_sn^b_t},\label{intersection}
\eea
where $W_{a/b}$ is the winding number of the stack $a/b$ around the 
whole $T^6$,
$L_{a/b}$ corresponds to the $U(1)$ line bundle associated to the 
magnetic
flux and $c_3(L_a\otimes L_b)$ is the third Chern class. The 
intersection number
$I_{ab}$ in (\ref{intersection}) is associated to the degeneracy of 
the 
Landau levels and therefore has to be integer. An obvious 
solution of this
requirement is to ask for the winding numbers $n^a_r$ of each stack 
$a$ to
satisfy\footnote{We thank R. Blumenhagen for useful communications on 
this point.}
\be
W_a =  n^a_rn^a_sn^a_t \,\,\,, \,\,\, \forall r,s,t \,\,{\rm 
with}\,\,  K_{rst}\neq 0.
\label{intersection2}
\ee
Since $I_{ab}$ depends only on the product of $W_a$ and $W_b$,
the above restriction is valid up to a sign ambiguity.
For each brane $a$,
there is a unique winding number $W_a$ around the whole torus $T^6$ 
which is given up to a sign by the product of winding numbers of
orthogonal 2-cycles. It corresponds to the geometrical picture where
the fundamental cycles of the torus are six 1-cycles and the
winding numbers $n_r$ around the fifteen different 2-cycles are not 
independent, but given in terms of products
of winding numbers around 1-cycles. Note that in this case, the 
7-brane
charge $q_t^a$ defined in (\ref{q7R}) reduces to $q_t^a 
=K_{rst}n_rn_sm_t$ without a sum over the indices $r,s$.

\subsection{R-R Moduli}\label{sec:rrmod}

We have seen above that under strong constraints on the magnetic 
fluxes, 
it is in principle possible to find $\mathcal{N}=1$ supersymmetric 
vacua in four 
dimensions with stabilized metric moduli. In sections
\ref{b&f}-\ref{explicit} and \ref{sec:Model-B}, 
we will give explicit examples where this is indeed achieved. Here, 
we 
want  to address the question of the remaining moduli. 
In the orientifold compactification 
$T^6/\mathbb{Z}_2$, apart from the metric and dilaton moduli, the 
four 
dimensional spectrum contains massless 2-forms, which arise in the 
R-R 
sector. They correspond to the 
internal components of the R-R 4-form $C_{(4)}$ which  survived the 
$\mathbb{Z}_{2}$-orientifold action defined in Section 
\ref{sec:SUSY2}. 
They are decomposed in elements of three different cohomology classes
$H^{1,1}(T^6)$, $H^{2,0}(T^6)$ and $H^{0,2}(T^6)$: 
\be
(C_{(4)})_{\mu\nu i \bar{j}} \quad ,\quad (C_{(4)})_{\mu\nu i j} 
,\quad (C_{(4)})_{\mu\nu \bar{i} \bar{j}}, \quad i,j=1,\dots,3
\label{rrmodC}\ee
where the indices $\mu,\nu$ refer to four dimensional spacetime : 
$\mu,\nu = 0,\dots,3$. The first nine 
2-forms $(C_{(4)})_{\mu\nu i \bar{j}}$ are dual to pseudo-scalars in 
four 
dimensions; they actually
form  linear multiplets with the K\"ahler moduli $J_{i\bar{j}}$. 
When the latter are fixed in the presence of magnetized fluxes, 
they give rise to  St\"uckelberg couplings that provide masses to 
some $U(1)$ gauge 
fields. 
This can be seen explicitly from the Wess Zumino action (\ref{WZ}) in 
ten dimensions: Consider  the gauge 
potential $A_M=(A_\mu,A_i)$ of a magnetized $U(1)$ with 
$A_k=-\frac{1}{2}F_{kl}u_l$. Its spacetime field 
strength $F_{(2)}=dA$ then couples to the 2-form $B_{(2)}^{i\bar{j}} 
= 
(C_{(4)})_{\mu\nu i \bar{j}}$ as:
\be
\int_{\mathcal{M}_{10}} C_4\wedge 
\mathcal{F}\wedge\mathcal{F}\wedge\mathcal{F} \rightarrow 
q_{i\bar{j}} 
\int_{\mathcal{M}_{4}}B_{(2)}^{i\bar{j}}\wedge F_{(2)},
\ee
where the couplings $q_{i\bar{j}}$ are  functions of the internal 
magnetic 
fluxes. As a result, some combination of 
the nine R-R 2-forms $(C_{(4)})_{\mu\nu i \bar{j}}$ is 
absorbed in the $U(1)$ gauge field which becomes massive.

The situation with the last six massless 2-forms  in (\ref{rrmodC}) 
is different. They are 
harmonic $(2,0)$ and $(0,2)$ forms on the internal torus and 
therefore elements of 
the cohomologies $H^{2,0}(T^6)$ and $H^{0,2}(T^6)$. By  contraction 
with the holomorphic 3-form $\Omega$ of $T^6$, we can  
construct from $(C_{(4)})_{\mu\nu i j}$ and $ 
(C_{(4)})_{\mu\nu\bar{i} 
\bar{j}}$ harmonic $(2,1)$ and $(1,2)$-forms on the torus:
\bea
B_{\mu\nu ij\bar{l}} &\sim& \Omega_{ij}^{\;\;\, 
\bar{k}}(C_{(4)})_{\mu\nu \bar{k} \bar{l}}\nonumber \\
B_{\mu\nu \bar{i}\bar{j}l} &\sim& \bar{ 
\Omega}_{\bar{i}\bar{j}}^{\;\;\, k}(C_{(4)})_{\mu\nu k l}\, .
\eea
To each harmonic $(2,0)$ and $(0,2)$ form, we can then associate a 
harmonic
$(2,1)$ and $(1,2)$-form, associated to the  complex structure 
moduli. Thus, the  nine elements of the complex structure  
$\tau_{ij}$ correspond to six purely (anti-) holomorphic 
metric moduli and three (anti-) holomorphic R-R moduli. 
As shown in \cite{Bianchi:2005yz}, the stabilization of the latter 
via the 
condition (\ref{csc1}) can be understood by a potential  generated 
through their mixing  with the NS-NS moduli.

\section{Closed string fluxes}\label{sec:3ff}
As argued in  section \ref{sec:moduli_stab}, all 
geometric moduli 
can be stabilized by turning on internal magnetic background 
fields. Moreover, the introduction of nine stacks of magnetized $D9$ 
branes 
can fix all complex structure and K\"ahler class moduli.
Furthermore, the R-R moduli complexifying the K\"ahler class 
are absorbed into the longitudinal degrees 
of freedom of the $U(1)$ gauge fields, which become massive. The 
remaining unfixed moduli correspond to the (complex) dilaton-axion 
field.

\subsection{Dilaton stabilization}

A possible stabilization mechanism for the dilaton is by turning on  
R-R and NS-NS 3-form closed string fluxes, that for generic 
Calabi-Yau 
compactifications can fix also the complex structure 
\cite{Frey:2002hf}. 
As we are going to combine the two mechanisms,  in this section we 
review 
briefly the main properties of 3-form fluxes. 

Let $H_{(3)}$ and $F_{(3)}$ be the field strengths of the NS-NS 
2-form 
$B_{(2)}$ and of the R-R 2-form $C_{(2)}$, respectively, 
$H_{(3)}=dB_{(2)}$
$F_{(3)}=dC_{(2)}$, subject as usual to the Dirac 
quantization 
condition in the compact space. In the basis $(\alpha_{a},\beta_{b})$ 
chosen in 
(\ref{H3basis}) of Appendix \ref{notations}, $H_{(3)}$ and $F_{(3)}$ 
can be written as 
\bea
\frac{1}{(2\pi)^{2}\alpha'}H_{(3)} & = & 
\sum_{a=0}^{h_{2,1}}\left(h^{a}_{1}\alpha_{a} + 
h^{a}_{2}\beta_{a}\right) \nonumber \\
\frac{1}{(2\pi)^{2}\alpha'}F_{(3)} & = &\sum_{a=0}^{h_{2,1}}\left( 
f^{a}_{1}\alpha_{a} + 
f^{a}_{2}\beta_{a}\right),
\label{hf_quanta}
\eea
where  $h_{1}^{a}$, $h_{2}^{a}$, $f_{1}^{a}$ and $f_{2}^{a}$ are 
integers. Using the complex dilaton modulus, one can then form the 
3-form $G_{(3)}$  
\be
 G_{(3)} = F_{(3)}- \phi H_{(3)} \quad , \quad \phi = 
C_{(0)} + i g_{s}^{-1}, \label{gfh} 
\ee
where $g_s$ is the string coupling.  The 3-form background fields 
preserve then a common supersymmetry with the 
$\mathbb{Z}_2$-orientifold projection of $T^6/\mathbb{Z}_2$  if the 
following  
conditions are fulfilled: $G_{(3)}$ has to be a primitive $(2,1)$ 
form \cite{Grana:2001xn}:
\be
 G_{(3)}\wedge J = 0 \quad , \quad  {G_{(3)}} \in H^{2,1}.
\label{primitivity}
\ee
Actually, the second of the conditions above corresponds to finding a 
minimum of the GVW 
superpotential \cite{Gukov:1999ya}
\be
W = \int_{T^{6}} G_{(3)} \wedge \Omega ,
\label{supot}
\ee
which then has to be covariantly constant with respect to all moduli, 
$D_IW = 0$,
or equivalently:
\be
W = 0 \quad , \quad \partial_{\phi}W = 0 \quad , \quad 
\partial_{\tau_{ij}}W = 0,
\label{supotcond}
\ee
where $\phi$ is defined in (\ref{gfh}).  Note that all primitive 
$(2,1)$-forms are 
imaginary self dual (ISD), $\star_6 G_{2,1} = i G_{2,1}$, where the 
star map $\star_6$ is the usual Hodge map on the torus. 

Let us analyze further the supersymmetry conditions 
(\ref{supotcond}). For given flux quanta (\ref{hf_quanta}), they can 
be understood as conditions on the dilaton and complex structure 
moduli. More precisely, using the symplectic structure 
(\ref{symp_st}), the superpotential (\ref{supot}) reads 
\be
W=\frac{1}{(2\pi)^{2}\alpha'} \int_{T^6} G_{(3)}\wedge \Omega  = 
-(f_{1}^{0}-\phi h_{1}^{0})\det\tau  + 
(f_{2}^{0}-\phi h_{2}^{0}) + (f_{1}^{ij}-\phi h_{1}^{ij})({\rm 
cof}\tau)_{ij} + (f_{2}^{ij}-\phi h_{2}^{ij})\tau_{ij}.
\ee
We can now express the three supersymmetry conditions 
(\ref{supotcond}) explicitly in the form :  
\bea
0 & = & -(f_{1}^{0}-\phi h_{1}^{0})det\tau  + 
(f_{2}^{0}-\phi h_{2}^{0}) + (f_{1}^{ij}-\phi h_{1}^{ij})({\rm 
cof}\tau)_{ij} + (f_{2}^{ij}-\phi h_{2}^{ij})\tau_{ij} 
\label{3ffcond1}\\
0 & = & h_{1}^{0}det\tau  - h_{2}^{0} - h_{1}^{ij}({\rm 
cof}\tau)_{ij} - h_{2}^{ij}\tau_{ij} \label{3ffcond2}\\
0 & = & -(f_{1}^{0}-\phi h_{1}^{0}) ({\rm cof}\tau)_{kl} + 
(f_{2}^{kl}-\phi h_{2}^{kl}) + (f_{1}^{ij}-\phi h_{1}^{ij}) 
\epsilon_{ikm}\epsilon_{jln}\tau_{mn}, \label{3ffcond3}
\eea
where ${\rm cof}\tau = (\det\tau)\tau^{-1,T}$. These are eleven 
conditions 
on the complex structure, 
parametrized by the nine elements $\tau_{ij}$ and the 
(complex) dilaton field $\phi$. It is then in principle 
possible to fix all 
complex structure and dilaton moduli in terms  of adequate 
quanta \cite{Frey:2002hf}. Let us now examine the 
primitivity condition $G_{(2,1)}\wedge J = 0$. We could naively think 
that this  can be interpreted, for given fluxes,  as 
conditions on the K\"ahler moduli. However, this condition is 
trivially satisfied in the case of generic Calabi-Yau 
compactifications, because 
there are no harmonic $(3,2)$ forms on these manifolds.  
Therefore, this condition can only become partially non-trivial  on 
 K\"ahler moduli for compactification  manifolds with 
more symmetries, such as the torus. 

There exist however alternative possibilities to fix the metric 
moduli.  As shown in section \ref{sec:SUSY}, the presence of 
internal magnetic fluxes leads to conditions on both the K\"ahler 
class
 (\ref{kc1a}) and complex structure moduli (\ref{M20_condition}). 
For generic Calabi-Yau spaces one can fix only the former, while
for toroidal compactifications it is possible to fix all metric 
moduli 
by a suitable choice of stacks of magnetized $D9$ branes. An explicit 
example
will be shown in section \ref{b&f}. On the other hand, the dilaton 
modulus remains unfixed, but can be stabilized using
3-form closed string fluxes. In fact, for fixed complex structure, 
the conditions 
(\ref{3ffcond1}), (\ref{3ffcond2}) and (\ref{3ffcond3}) constrain 
exclusively the dilaton. Moreover, as the K\"ahler form is fixed 
by the presence of magnetic fields, 
the primitivity condition $G_{(2,1)}\wedge J = 0$ restricts 
the possible fluxes $G_{(2,1)}$ we can switch on. Finally, the value 
of the string coupling we can obtain in this way is strongly 
constrained by the tadpole conditions. 
The latter can be read off from the topological coupling of the 
3-form 
fluxes with the R-R 4-form $C_{(4)}$ potential in the 
effective action of the ten-dimensional type IIB supergravity:
\be
S_{CS} = \frac{1}{4i (2\pi)^7 \alpha'^4}\int_{\mathcal{M}_{10}} 
\frac{C_{(4)} \wedge G_{(3)}\wedge \bar{G}_{(3)} }{{\rm Im}\phi} = 
-\mu_3 \, \frac{1}{2} {1\over (2\pi)^4 
{\alpha'}^2}\int_{\mathcal{M}_{10}}C_{(4)}\wedge H_{(3)}\wedge 
F_{(3)},
\label{IIBCS}
\ee 
where we defined the R-R charge $\mu_3$ in terms of $\alpha'$ as 
$\mu_3 = 
(2\pi)^{-3}{\alpha'}^{-2}$.
The coupling to $C_{(4)}$ of the magnetized $D9$ branes 
is given in (\ref{WZ}), while the coupling of the $O3$ orientifold 
plane reads
\be
S_{O3} = \mu_3 Q_{O3} \int_{\mathcal{M}_4}C_{(4)}\, ,
\label{EffActO3}
\ee
where the charge $Q_{O3}$ of $O3$ planes has  been defined in section 
\ref{sec:tadpoles}. Therefore, the integrated Bianchi identity for 
the modified R-R 5-form field strength $F_{(5)}$ reads
\be
 -\frac{1}{2}\frac{1}{(2\pi)^{4}\alpha'^{2}}\int_{T^{6}}H_{(3)}\wedge 
F_{(3)} + q_{3,R} + Q_{O3} = 0,
\label{N3}
\ee
 where the factor $1\over 2$ comes from the fact that the volume of 
the orientifold 
 $T^6/\mathbb{Z}_2$ is half the volume of the torus 
$T^6$.\footnote{Note that it does not come from the factor 
 ${1\over 2}$ in (\ref{IIBCS}) which is compensated by the magnetic 
coupling 
 to $C_{(4)}$; see \cite{Giddings:2001yu} for more details. }

It follows from the ISD condition, that the contribution to (\ref{N3})
coming from the 3-form flux is always positive :
\be
N_3 
=:-\frac{1}{2}\frac{1}{(2\pi)^{4}\alpha'^{2}}\int_{T^{6}}H_{(3)}\wedge 
F_{(3)}= \frac{1}{2g_s} \int_{T^6} H_{(3)}\wedge \star_6 H_{(3)}>0.
\label{N3pos}
\ee
The second source for 3-brane charges in (\ref{N3}) comes from the 
internal magnetic fluxes. As shown in  section \ref{sec:tadpoles}, 
each stack of magnetized $D9$ branes with magnetic fluxes 
switched on in three orthogonal directions of $T^6$ contributes 
positively to the 3-brane charge (\ref{q3R}). Finally, the 3-brane
tadpole could also receive contributions from ordinary $D3$ branes. 
All together, 
the tadpole condition (\ref{t3}) is now modified as 
\be
N_3 + q_{3,R} + N_{D3} + Q_{O3} = 0,
\label{tad3ff}
\ee
where $Q_{O3} = -16$. As the first three terms in the l.h.s. of 
equation
(\ref{tad3ff}) contribute positively, the possible values of  $N_3$ 
as well of $q_{3,R}$ are  bounded. This restricts strongly the 
possible 
values of the string coupling $g_s$. Since the tadpole condition 
(\ref{tad3ff}) 
asks for $N_3$ to  remain of order one, the only possibility for the 
string coupling 
to be fixed at a small value is to get a large contribution from the 
integral 
$\int_{T^6}H_{(3)}\wedge \star_6 H_{(3)}$. This depends on the quanta 
$h_1^a$ , $h_2^a$ of (\ref{hf_quanta}) and on the Hodge star 
operator. 
The latter only depends on the complex structure 
\cite{Ceresole:1995ca}. It is therefore in principle possible to fix 
the string coupling $g_s$ at small value and to keep the contribution 
$N_3$ at fixed value by stabilizing the integral 
$\int_{T^6}H_{(3)}\wedge \star_6 H_{(3)}$ at large value with the 
help of either internal magnetic fields or 3-form fluxes. This will 
be discussed in more details in section \ref{sec:led}.

\subsection{Quantized NS-NS B field}\label{sec:NSNSB}

Further restrictions on  fluxes arise from the quantization condition 
in the orientifold 
$T^{6}/\mathbb{Z}_{2}$, as compared to the torus. As explained in 
\cite{Frey:2002hf}, the quanta of  NS-NS and R-R 3-form fluxes have 
to be even along any 3-cycle of $T^{6}/\mathbb{Z}_{2}$. This  
remains  valid in the presence of magnetic fluxes, as well. However, 
the situation changes if one introduces a 
non-trivial NS-NS B field in some of the 2-cycles of the torus. 
Consider for instance the case where the B field is switched on only 
in one  
2-cycle, say $[x_3y_3]$: $B_{x_3y_3} = \frac{\alpha'}{(2\pi)^2}b$, 
where $b=0$ or $1/2$. Its consequences are:
\begin{itemize}
\item
A change in the spectrum of the 
open string sector \cite{Bquant}. The first Chern number  
$m^a_{x_3y_3}$ of the magnetic fluxes of all stacks $a=1,\cdots , K$ 
gets shifted to 
$\tilde{m}_{x_3y_3} = m_{x_3y_3}^a +b n_{x_3y_3}^a$.

\begin{figure}[tbp]
    \centering
    \includegraphics[height=6cm]{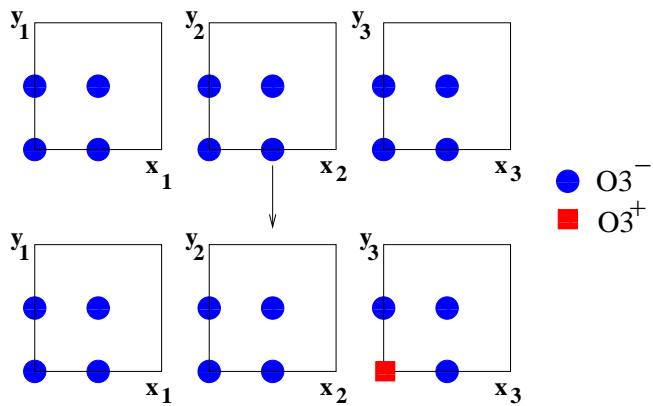}
    \caption{$O3$ plane configuration in case of discrete torsion in 
the direction $[x_3y_3]$.}
    \label{fig:b_o}
\end{figure}

\item A modification of  the 
configuration of $O3$-planes. In the orientifold compactification 
$T^6/\mathbb{Z}_2$, there are 64 fixed points 
 where the different 
$O3$ planes sit. All of them have  negative tension and charge. 
However, 
for $b=1/2$, 16 of the 64 $O3$ planes become of the type 
$O3^+$, which have positive tension and charges \cite{Witten:1997bs}. 
The remaining 48 are of the usual type $O3\equiv O3^-$.
The 3-brane tadpole condition 
is therefore modified. As in our conventions each $O3$ orientifold 
plane
carries $1/4$ unit of (negative) charge, the total charge 
contribution of 
the different orientifold planes for $b=1/2$ is not anymore $-16$  
but~$-8$:
\be
N_{O3^+} - N_{O3^-} = \frac{1}{4}( 16 - 48) = - 8.
\ee
The tadpole condition (\ref{tad3ff}) is then modified to
\be
N_3 + \tilde{q}_{3,R} + N_{D3} = 8 \quad , \quad \tilde{q}_{3,R} 
=\sum_{a=1}^{K}\,\, N_a \sum_{r,s,t}\,\, 
\mathcal{K}_{rst}\tilde{m}^a_r 
m^a_s m^a_t \, .
\label{t3NSNSB}
\ee
In the modified 3-brane charge $\tilde{q}_{3,R}$ induced by the 
magnetic fields, it is implicitly 
assumed that the only shifted Chern numbers $\tilde{m}^a_r$ 
correspond to the 2-cycle 
carrying the B-field; in our example, it is $\tilde{m}^a_{x_3y_3}$.

\item
A modification of 
the quantization condition for the NS-NS 3-form fluxes $H_{(3)}$  
\cite{Frey:2002hf}. Consider first a NS-NS 3-form  switched on in a 
3-cycle $\gamma$ of the torus $T^6$. If $\gamma$ crosses 
an odd number of orientifold planes of the type $O3^+$, the 
corresponding quanta $h_\gamma$ have to be odd integers, while 
when the crossing number  is even, $h_\gamma$ has to be even. Let us 
consider now the case where $b=1/2$. As depicted in Figure 
\ref{fig:b_q}, the sixteen $O3^+$ planes are located at one of the 
four fixed points of the third torus $[x_3y_3]$. We can easily see 
that 
the only 3-cycles of $T^6$, whose crossing number with the $O3^+$ 
planes is odd are the following ones:
\be
[x_iy_ix_3] \quad , \quad [x_iy_iy_3] \quad , \quad i=1,2 \,\,\, .
\ee
They correspond to 3 cycles wrapping a `diagonal' 2-cycle 
$[x_iy_i]$ as well as one of the 1-cycles $x_3$ or $y_3$. 
They are located at one fixed point of the last 2-torus. As a result, 
the following quanta (\ref{hf_quanta}) of $H_{(3)}$ can be odd:
\be
h_1^{12} \quad , \quad h_1^{21}\quad , \quad  h_2^{12} \quad , \quad 
h_2^{21}\, .
\label{oddquanta}
\ee

\begin{figure}[tbp]
    \centering
    \includegraphics[height=3cm]{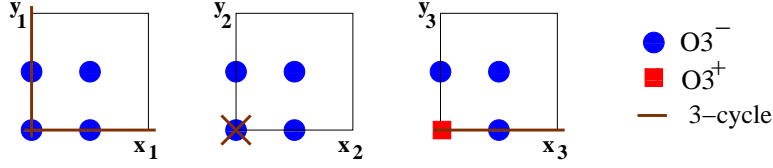}
    \caption{Example of 3-cycle with an odd crossing number of 
$O3^+$'s. The cycle $[x_1y_1x_3]$ crosses a single $O3^+$.  }
    \label{fig:b_q}
\end{figure}

\end{itemize}

\section{Branes and fluxes}\label{b&f}

In this section we present the different stacks of magnetized $D9$ 
branes we need in order to satisfy the supersymmetry conditions
(\ref{M20_condition}), (\ref{kc1a}), the positivity requirements 
(\ref{no_anti}), (\ref{no_anti2}) and (\ref{Jcond}), and the tadpole 
cancellations (\ref{t7}) 
and (\ref{t3}). For the shake of simplicity, our aim is to stabilize 
the 
moduli to a geometry of a factorized torus
$T^6$ as $T^2\times T^2\times T^2$. This implies in particular that
the off-diagonal components of the complex structure, defined 
in terms of the real coordinates $x^i$, $y^i$ ($i=1,2,3$)
through eq.~(\ref{complex_structure}), should vanish.

 The constraints on the complex structure matrix $\tau^{ij}$ are
derived from eq.~(\ref{M20_condition}). We notice that, in 
order to have the off-diagonal components of the 
complex structure moduli vanishing, one needs to turn on 
certain off-diagonal components of magnetic 
fluxes on the $D9$ branes. They are characterized by 
rational numbers $p^a$'s, defined as the ratios of the
quantum numbers $m$ and $n$  given in eq.~(\ref{diraca}). 
These fluxes turn out to be of the type $p^a_{x^iy^j}$,  
$p^a_{x^ix^j}$ and $p^a_{y^iy^j}$, with $i\neq j$. 
However, we will find out later that
 off-diagonal fluxes of these types have to be necessarily  
accompanied by certain diagonal fluxes of the type
$p^a_{x^iy^i}$ $(i=1,2,3)$, as well. Taking these restrictions into
account,  the following non-zero fluxes are turned on along the
branes in stack-1:
\be
1.\;\; [p^1_{x^1y^2},\;\; p^1_{x^2y^1},\;\; p^1_{x^1y^1},\;\; 
p^1_{x^3y^3}]
                 \neq 0, 
\label{no.1}
\ee
with the remaining  components of the flux being set to zero, by 
choosing the corresponding 
Chern numbers $m=0$ in eq.~(\ref{diraca}). 
The windings $n_r$ can be zero along some of 
the 2-cycles $C_{(2)}^r$, even if the corresponding magnetic flux 
vanishes. 
However, since the magnetized branes are $D9$'s, they have to cover 
the whole 
internal space $T^6/\mathbb{Z}_2$. This means that the effective 
winding number 
$K_{rst}n^a_rn^a_sn^a_t$
around (the 6-cycle of) $T^6/\mathbb{Z}_2$ has to be non-zero.

Similarly to (\ref{no.1}), we choose for stack-2:
\be
2.\;\; [p^2_{x^2y^3},\;\; p^2_{x^3y^2},\;\; p^2_{x^2y^2},\;\; 
p^2_{x^1y^1}]
                 \neq 0
\label{no.2}
\ee
and for stack-3:
\be
3.\;\; [p^3_{x^3y^1},\;\; p^3_{x^1y^3},\;\; p^3_{x^3y^3},\;\; 
p^3_{x^2y^2}]
                 \neq 0\, . 
\label{no.3}
\ee
As we will see below, the supersymmetry condition 
(\ref{M20_condition}) 
on the stacks of branes 1-3, with fluxes turned on according to  
eqs.~(\ref{no.1})-(\ref{no.3}), imply that all off-diagonal 
components of 
$\tau^{ij}$ ($i\neq j$) are set to zero. Moreover, these conditions 
fix the ratios of the diagonal components $\tau^{ii}$ in terms of 
the ratios of $p^a_{x^iy^j}$ in the different brane stacks. We will 
also 
show that some magnetized branes will  play a role in setting three 
independent 
combinations of the off-diagonal components of the K\"ahler class 
moduli
$J_{i\bar{j}}$ to zero. In order to get similar conditions on the 
remaining off-diagonal components of the K\"ahler moduli, we  
introduce 
three more brane stacks with the following non-vanishing flux 
components:
\be
4.\;\; [p^4_{x^1x^2},\;\; p^4_{y^1y^2},\;\; p^4_{x^1y^1},\;\; 
p^4_{x^3y^3}]
                 \neq 0, 
\label{no.4}
\ee
\be
5.\;\; [p^5_{x^2x^3},\;\; p^5_{y^2y^3},\;\; p^5_{x^2y^2},\;\; 
p^5_{x^1y^1}]
                 \neq 0, 
\label{no.5}
\ee
and
\be
6.\;\; [p^6_{x^3x^1},\;\; p^6_{y^3y^1},\;\; p^6_{x^3y^3},\;\; 
p^6_{x^2y^2}]
                 \neq 0.
\label{no.6}
\ee

Studying various possibilities of string 
constructions 
incorporating  moduli stabilization, we will also introduce in some 
cases  
six more copies of brane stacks, called stack-$1^\prime$ - 
stack-$6^\prime$. 
These branes have the same diagonal fluxes (and with brane 
multiplicities 
$N_{a'}=N_a$) as their unprimed 
counterparts, but  off-diagonal components with opposite 
sign:
\be
1'.\;\; [- p^1_{x^1y^2},\;\; - p^1_{x^2y^1},\;\; p^1_{x^1y^1},\;\; 
           p^1_{x^3y^3}]  \neq 0, 
\label{no.1'}
\ee
\be
2'.\;\; [-p^2_{x^2y^3},\;\; -p^2_{x^3y^2},\;\; p^2_{x^2y^2},\;\; 
          p^2_{x^1y^1}]  \neq 0, 
\label{no.2'}
\ee
\be
3'.\;\; [-p^3_{x^3y^1},\;\; - p^3_{x^1y^3},\;\; p^3_{x^3y^3},\;\; 
          p^3_{x^2y^2}]  \neq 0, 
\label{no.3'}
\ee
\be
4'.\;\; [-p^4_{x^1x^2},\;\; - p^4_{y^1y^2},\;\; p^4_{x^1y^1},\;\; 
          p^4_{x^3y^3}] \neq 0, 
\label{no.4'}
\ee
\be
5'.\;\; [- p^5_{x^2x^3},\;\; - p^5_{y^2y^3},\;\; p^5_{x^2y^2},\;\; 
             p^5_{x^1y^1}] \neq 0, 
\label{no.5'}
\ee
\be
6'.\;\; [- p^6_{x^3x^1},\;\; - p^6_{y^3y^1},\;\; p^6_{x^3y^3},\;\; 
           p^6_{x^2y^2}] \neq 0. 
\label{no.6'}
\ee

The stacks 1-6 (or alternatively stacks $1^\prime$-$6^\prime$), when 
used 
with some other branes with diagonal fluxes along $[x^i y^i]$ (called 
stacks 7-9), 
give six independent conditions on the K\"ahler moduli 
$J_{i\bar{j}}$, ($i\neq j$) and force them to vanish. 
In our examples, we choose the stacks 7-9 
having only two non-zero diagonal components of  magnetic fluxes. 
The magnetic fields along these branes
are required to satisfy 
the consistency conditions mentioned in section \ref{sec:tadpoles}
and are sufficient to fix all
diagonal components of the K\"ahler moduli $J$, as well.
More precisely,  the fluxes in stacks 7-9 read:
\be
7.\;\; [ p^7_{x^1y^1}  \neq 0,\;\; p^7_{x^2y^2}  = 0,\;\;  
        p^7_{x^3y^3} \neq 0],  
\label{no.7}
\ee
\be
8.\;\; [ p^8_{x^1y^1}  \neq 0,\;\; p^8_{x^2y^2}  \neq 0,\;\;  
        p^8_{x^3y^3} = 0],
\label{no.8}
\ee
\be
9.\;\; [ p^9_{x^1y^1} = 0,\;\; p^9_{x^2y^2}  \neq 0.\;\;  
        p^9_{x^3y^3} \neq 0].
\label{no.9}
\ee

Another possibility to satisfy 
the consistency conditions mentioned in section \ref{sec:tadpoles} is 
to introduce some stacks of
branes with off-diagonal fluxes which do not contribute to the 
3-brane tadpole:
\be
10.\;\; [ p^{10}_{x^1y^1}  \;,\; p^{10}_{x^3y^3}  \;,\;  
p^{10}_{x^1y^3}\;, \, p^{10}_{x^3y^1}]\neq 0,\,\, \, \,\, 
10'.\;\; [ p^{10'}_{x^1y^1}  \;,\; p^{10'}_{x^3y^3}  \;,\;  
p^{10'}_{x^1x^3}\;,\;  p^{10'}_{y^3y^1}]\neq 0, 
\label{no.7''}
\ee
\be
11.\;\; [ p^{11}_{x^1y^1}  \;,\; p^{11}_{x^2y^2}  \;,\;  
p^{11}_{x^1y^2}\;,\;  p^{11}_{x^2y^1}]\neq 0,\,\, \, \,\,
11'.\;\; [ p^{11'}_{x^1y^1}  \;,\; p^{11'}_{x^2y^2}  \;,\;  
p^{11'}_{x^1x^2}\;,\;  p^{11'}_{y^2y^1}]\neq 0,
\label{no.8''}
\ee
\be
12.\;\; [ p^{12}_{x^2y^2}  \;,\; p^{12}_{x^3y^3}  \;,\;  
p^{12}_{x^2y^3}\;,\;  p^{12}_{x^3y^2}]\neq 0,\,\, \,\, \,\, 
12'.\;\; [ p^{12'}_{x^2y^2}  \;,\; p^{12'}_{x^3y^3}  \;,\;  
p^{12'}_{x^2x^3}\;,\;  p^{12'}_{y^3y^2}] \neq 0.
\label{no.9''}
\ee
Of course, one  has also the possibility of introducing branes with 
non-zero fluxes along all  diagonal elements. Such branes are,
however, not used in the examples we present below, for simplicity 
and for minimizing the 3-brane tadpole contribution. 

We are now ready to examine the moduli stabilization when 
different combinations of branes, mentioned above, are used.

\section{Explicit construction: Model-A with $q_{3,R} = 
6$}\label{explicit}

In this section, we analyze the conditions 
(\ref{M20_condition}), (\ref{kc1a}), (\ref{Jcond}),
(\ref{t7}), (\ref{t3}),(\ref{no_anti}) and (\ref{no_anti2}) in more 
detail and present explicit examples
when the twelve brane stacks 1-6, 10-12 and $10'$-$12'$ are used. 
We first discuss 
complex structure moduli stabilization, and next, in 
subsections 
\ref{subsection-kahler}-\ref{subsection-explicit-flux+moduli}, we 
show the 
stabilization of the K\"ahler class moduli, as well. 
These branes together contribute $q_{3,R} = 6$ to the 
3-brane tadpoles;
tadpole cancellation will be discussed in 
subsection \ref{subsection-tadpole-A}.

\subsection{Stabilization of complex structure moduli}\label{model-A}

We show below that all  complex structure moduli are stabilized 
using only the stacks of branes 1-6, with magnetic fluxes given in 
eqs.~(\ref{no.1})-(\ref{no.6}). In fact the situation remains 
similar to the (T-dual) case studied in \cite{AM}, and we only give 
the final 
conditions following from the vanishing of the $F_{(2, 0)}$ components
({\it c.f.} (\ref{csc1})), as given in (\ref{M20_condition}).

First, the brane stacks 1-3 restrict the off-diagonal elements of 
the complex 
structure matrix by
a set of six linear equations for the six variables, 
$\tau^{12}$, $\tau^{23}$, $\tau^{31}$, $\tau^{21}$, $\tau^{32}$, 
$\tau^{13}$:
\bea
\begin{pmatrix} 0 &  p^1_{x^2y^1} & - p^1_{x^3y^3}  \cr
\cr
- p^2_{x^1y^1} & 0 & p^2_{x^3y^2}  \cr               
\cr
p^3_{x^1y^3} & - p^3_{x^2y^2} & 0 
\end{pmatrix}
\begin{pmatrix}\tau^{12}\cr \cr \tau^{23} \cr \cr 
\tau^{31}\end{pmatrix} = 
\begin{pmatrix}
-p^1_{x^1y^1}\tau^{13}\cr \cr -p^2_{x^2y^2}\tau^{21}\cr\cr 
-p^3_{x^3y^3}\tau^{32}
\end{pmatrix},
\label{complex-matrix1}
\eea
\bea
\begin{pmatrix} 
 - p^1_{x^1y^2} & 0  & p^1_{x^3y^3} \cr
\cr               
 0  &  p^3_{x^2y^2} & - p^3_{x^3y^1} \cr
\cr
  p^2_{x^1y^1} & -p^2_{x^2y^3} & 0 \end{pmatrix}
\begin{pmatrix} \tau^{13}\cr\cr \tau^{21}\cr\cr 
\tau^{32}\end{pmatrix} = 0.
\label{complex-matrix2}
\eea
As we will see later on, for the specific values of  magnetic
fluxes that we turn on along the branes,
the matrix appearing in eq.~(\ref{complex-matrix2})
turns out to be singular and implies the equality:
\be
\tau^{13} = \tau^{21} = \tau^{32}.
\label{tau-0-2}
\ee
Moreover, the matrix appearing in the l.h.s. of 
eq.~(\ref{complex-matrix1})
is also singular, and using the result (\ref{tau-0-2}), one obtains:
\be
\tau^{13} = 0\quad ;\quad\tau^{12} = \tau^{23} = \tau^{31}.
\label{tau-0-1}
\ee
Finally, using the constraint (\ref{M20_condition}) for one of the 
branes
4, 5 or 6, one obtains that all off-diagonal components 
of the complex-structure vanish:
\be
\tau^{12} = \tau^{13} = \tau^{21} =  \tau^{23}
        = \tau^{31} = \tau^{32} = 0\, .
\label{tau-0}
\ee

The brane stacks 1-6 also restrict the diagonal elements of the 
matrix 
$\tau$, and they satisfy the following conditions:
\be
 \frac{\tau^{11}}{\tau^{22}} = \frac{p^1_{x^2y^1}}{ p^1_{x^1y^2}} 
                                \equiv K_1,\;\;\;
\frac{\tau^{22}}{\tau^{33}} =  \frac{p^2_{x^3y^2}}{p^2_{x^2y^3}}
                                \equiv K_2,\;\;\;
\frac{\tau^{33}}{\tau^{11}} = \frac{p^3_{x^1y^3}} {p^3_{x^3y^1}}
                                \equiv K_3,
\label{ratio}
\ee
and 
\be
 \tau^{11} \tau^{22} = - \frac{p^4_{y^1y^2}}{ p^4_{x^1x^2}} 
                                \equiv - K_4,\;\;\;
\tau^{22} \tau^{33} =  - \frac{p^5_{y^2y^3}}{p^5_{x^2x^3}}
                                \equiv - K_5,\;\;\;
\tau^{33} \tau^{11} =  - \frac{p^6_{y^3y^1}}{p^6_{x^3x^1}}
                                \equiv - K_6.
\label{product}
\ee
Following \cite{AM}, we use $K_1$, $K_3$ and $K_4$ as independent 
parameters. Consistency between
eqs.~(\ref{ratio}) and (\ref{product}) then implies:
\be
K_2 = {1\over K_1 K_3},\;\;K_5 = {K_3 K_4},\;\;
K_6 = {K_1 K_3 K_4}.\;\;
\label{consistency}
\ee
Since we will look for solutions where $\tau^{ii}$ are all 
purely imaginary, this further imposes a positivity condition on 
$K_i$'s:
\be
  K_i > 0,\;\; {\rm for} \;\; i=1,..,6. 
\label{ki-positive}
\ee
The solutions for the diagonal elements $\tau^{ii}$ ($i=1,2,3$)
are then given by:
\be
\tau^{11} = i\sqrt{K_1K_4},\;\;\;\tau^{22} = i\sqrt{K_4\over 
K_1},\;\;\;
\tau^{33} = i\sqrt{K_1K_4} K_3.\;\;\;
\label{diagonal-tau}
\ee
We have therefore determined the complex structure moduli completely,
given by the equations (\ref{tau-0}) and (\ref{diagonal-tau}).
Using this form of the complex structure, it can also be  
easily verified that the stacks of branes 7-9, having fluxes only 
along 
diagonals $x^iy^i$, do not impose
any further constraints on it. We go on now to the stabilization of  
the K\"ahler class moduli.

\subsection{Stabilization of K\"ahler class moduli: 
constraints on fluxes}\label{subsection-kahler}

In this subsection we derive the constraints on  
magnetic fluxes, for the stack of branes 1-6, 
10-12 and $10'$-$12'$, defined
in section \ref{b&f}, in order to 
obtain the stabilization of the K\"ahler class moduli.
For this purpose, we analyze the supersymmetry condition 
(\ref{kc1a}) for these stacks. 
As the complex structure has been 
stabilized to the diagonal form $\tau_{ij} = i \delta_{ij}$, the flux 
content given earlier in eqs.~(\ref{no.1})-(\ref{no.3})
can be written with the help of (\ref{11part}) as 
\be
1.\;\; [ \mathcal{F}^1_{1\bar{2}} = \mathcal{F}^1_{2\bar{1}},\;\; 
\mathcal{F}^1_{1\bar{1}},\;\; 
           \mathcal{F}^1_{3\bar{3}}]  \neq 0, 
\label{no.1-complex}
\ee
\be
2.\;\; [ \mathcal{F}^2_{2\bar{3}} = \mathcal{F}^2_{3\bar{2}},\;\; 
\mathcal{F}^2_{2\bar{2}},\;\; 
           \mathcal{F}^2_{1\bar{1}}]  \neq 0, 
\label{no.2-complex}
\ee
\be
3.\;\; [ \mathcal{F}^3_{3\bar{1}} = \mathcal{F}^3_{1\bar{3}},\;\; 
\mathcal{F}^3_{3\bar{3}},\;\; 
           \mathcal{F}^3_{2\bar{2}}]  \neq 0. 
\label{no.3-complex}
\ee 
In these expressions for fluxes, we have used complex coordinates, 
instead of the real ones used in eqs.~(\ref{no.1})-(\ref{no.3}), 
related to each other through eqs.~(\ref{complex_structure}) and 
(\ref{diraca}). In particular, the 
off-diagonal components of the fluxes in eqs.~(\ref{no.1-complex}) - 
(\ref{no.3-complex})
are purely imaginary. On the other hand, the off-diagonal fluxes in 
 stacks 4-6 are 
real and they have the form:
\be
4.\;\; [\mathcal{F}^4_{1\bar{2}} =  - \mathcal{F}^4_{2\bar{1}},\;\; 
\mathcal{F}^4_{1\bar{1}},\;\; 
          \mathcal{F}^4_{3\bar{3}}] \neq 0, 
\label{no.4-complex}
\ee
\be
5.\;\; [\mathcal{F}^5_{2\bar{3}} = - \mathcal{F}^5_{3\bar{2}},\;\; 
\mathcal{F}^5_{2\bar{2}},\;\; 
             \mathcal{F}^5_{1\bar{1}}] \neq 0, 
\label{no.5-complex}
\ee
\be
6.\;\; [ \mathcal{F}^6_{3\bar{1}} = - \mathcal{F}^6_{1\bar{3}},\;\; 
\mathcal{F}^6_{3\bar{3}},\;\; 
           \mathcal{F}^6_{2\bar{2}}] \neq 0.
\label{no.6-complex}
\ee 
For the remaining branes 10-12 and $10'$-$12'$, the non-vanishing 
diagonal flux 
components in complex coordinates are:
\be
10.\;\; [\mathcal{F}^{10}_{1\bar{1}},\;\; 
\mathcal{F}^{10}_{3\bar{3}}, \;\;\mathcal{F}^{10}_{1\bar{3}}
= \mathcal{F}^{10}_{3\bar{1}}] \neq 
0,\;\;\;
10'.\;\; [\mathcal{F}^{10'}_{1\bar{1}},\;\; 
\mathcal{F}^{10'}_{3\bar{3}}, \;\;\mathcal{F}^{10'}_{1\bar{3}}
= - \mathcal{F}^{10'}_{3\bar{1}}] \neq 
0, 
\label{no.7-complex}
\ee
\be
11.\;\; [\mathcal{F}^{11}_{2\bar{2}},\;\; \mathcal{F}^{11}_{1\bar{1}} 
,\;\; \mathcal{F}^{11}_{1\bar{2}} 
= \mathcal{F}^{11}_{2\bar{1}}] \neq 
0,\;\;\;
11'.\;\; [\mathcal{F}^{11'}_{2\bar{2}},\;\; \mathcal{F}^{11'}_{1\bar{1}} 
,\;\; \mathcal{F}^{11'}_{1\bar{2}}
= - \mathcal{F}^{11'}_{2\bar{1}}] \neq 
0, 
\label{no.8-complex}
\ee
\be
12.\;\; [\mathcal{F}^{12}_{3\bar{3}},\;\; 
\mathcal{F}^{12}_{2\bar{2}},\;\; \mathcal{F}^{12}_{2\bar{3}}
= \mathcal{F}^{12}_{3\bar{2}}] \neq 
0,\;\;\;
12'.\;\; [\mathcal{F}^{12'}_{3\bar{3}},\;\; 
\mathcal{F}^{12'}_{2\bar{2}},\;\; \mathcal{F}^{12'}_{2\bar{3}}
= - \mathcal{F}^{12'}_{3\bar{2}}] \neq 
0.\;\;\;
\label{no.9-complex}
\ee

We now analyze the supersymmetry condition (\ref{kc1a})
and find that it puts several restrictions on the fluxes that are 
turned on. 
Expressing eq.~(\ref{kc1a}) in components, we obtain for  brane-1:
\be
 (J\wedge J\wedge J)_{1\bar{1}2\bar{2}3\bar{3}} = 
     J_{2\bar{2}}\mathcal{F}^1_{1\bar{1}}\mathcal{F}^1_{3\bar{3}}
                   - 
J_{3\bar{3}}\mathcal{F}^1_{1\bar{2}}\mathcal{F}^1_{2\bar{1}}
                - ( J_{1\bar{2}} + J_{2\bar{1}}) 
                  \mathcal{F}^1_{3\bar{3}}\mathcal{F}^1_{2\bar{1}},
\label{kahler-1}
\ee
where in writing the last term we have also made use of the 
condition   
$\mathcal{F}^1_{2\bar{1}} = \mathcal{F}^1_{1\bar{2}}$ given in 
eq.~(\ref{no.1-complex}). Similarly, we have for brane-2 and 
brane-3:  
\be
 (J\wedge J\wedge J)_{1\bar{1}2\bar{2}3\bar{3}} 
                = 
J_{3\bar{3}}\mathcal{F}^2_{2\bar{2}}\mathcal{F}^2_{1\bar{1}}
                   - 
J_{1\bar{1}}\mathcal{F}^2_{2\bar{3}}\mathcal{F}^2_{3\bar{2}}
                - ( J_{2\bar{3}} + J_{3\bar{2}} ) 
                \mathcal{F}^2_{1\bar{1}}\mathcal{F}^2_{3\bar{2}},
\label{kahler-2}
\ee
\be
 (J\wedge J\wedge J)_{1\bar{1}2\bar{2}3\bar{3}} 
                = 
J_{1\bar{1}}\mathcal{F}^3_{3\bar{3}}\mathcal{F}^3_{2\bar{2}}
                   - 
J_{2\bar{2}}\mathcal{F}^3_{3\bar{1}}\mathcal{F}^3_{1\bar{3}}
                - (J_{3\bar{1}} + J_{1\bar{3}}) 
                \mathcal{F}^3_{2\bar{2}}\mathcal{F}^3_{1\bar{3}}.
\label{kahler-3}
\ee
For branes 4-6, on the other hand, we get:
\be
 (J\wedge J\wedge J)_{1\bar{1}2\bar{2}3\bar{3}} 
                = 
J_{2\bar{2}}\mathcal{F}^4_{1\bar{1}}\mathcal{F}^4_{3\bar{3}}
                   - 
J_{3\bar{3}}\mathcal{F}^4_{1\bar{2}}\mathcal{F}^4_{2\bar{1}}
                - ( J_{1\bar{2}} - J_{2\bar{1}}) 
                  \mathcal{F}^4_{3\bar{3}}\mathcal{F}^4_{2\bar{1}},
\label{kahler-4}
\ee
\be
 (J\wedge J\wedge J)_{1\bar{1}2\bar{2}3\bar{3}} 
                = 
J_{3\bar{3}}\mathcal{F}^5_{2\bar{2}}\mathcal{F}^5_{1\bar{1}}
                   - 
J_{1\bar{1}}\mathcal{F}^5_{2\bar{3}}\mathcal{F}^5_{3\bar{2}}
                - ( J_{2\bar{3}} - J_{3\bar{2}} ) 
                \mathcal{F}^5_{1\bar{1}}\mathcal{F}^5_{3\bar{2}},
\label{kahler-5}
\ee
\be
 (J\wedge J\wedge J)_{1\bar{1}2\bar{2}3\bar{3}} 
                = 
J_{1\bar{1}}\mathcal{F}^6_{3\bar{3}}\mathcal{F}^6_{2\bar{2}}
                   - 
J_{2\bar{2}}\mathcal{F}^6_{3\bar{1}}\mathcal{F}^6_{1\bar{3}}
                - (J_{3\bar{1}} - J_{1\bar{3}}) 
                \mathcal{F}^6_{2\bar{2}}\mathcal{F}^6_{1\bar{3}}.
\label{kahler-6}
\ee
Finally, the supersymmetry condition for branes 10-12
and $10'$-$12'$  implies:
\be
  (J\wedge J\wedge J)_{1\bar{1}2\bar{2}3\bar{3}} 
                = J_{2\bar{2}} ( \mathcal{F}^{10}_{3\bar{3}} 
\mathcal{F}^{10}_{1\bar{1}}-|\mathcal{F}^{10}_{1\bar{3}}|^2 )
= J_{2\bar{2}} ( \mathcal{F}^{10'}_{3\bar{3}} 
\mathcal{F}^{10'}_{1\bar{1}}-|\mathcal{F}^{10'}_{1\bar{3}}|^2 ),
\label{kahler-7}
\ee
\be
  (J\wedge J\wedge J)_{1\bar{1}2\bar{2}3\bar{3}} 
                = J_{3\bar{3}} ( \mathcal{F}^{11}_{1\bar{1}} 
\mathcal{F}^{11}_{2\bar{2}}-|\mathcal{F}^{11}_{1\bar{2}}|^2 )
= J_{3\bar{3}} ( \mathcal{F}^{11'}_{1\bar{1}} 
\mathcal{F}^{11'}_{2\bar{2}}-|\mathcal{F}^{11'}_{1\bar{2}}|^2 ),
\label{kahler-8}
\ee
\be
  (J\wedge J\wedge J)_{1\bar{1}2\bar{2}3\bar{3}} 
= J_{1\bar{1}} ( \mathcal{F}^{12}_{3\bar{3}} 
\mathcal{F}^{12}_{2\bar{2}}-|\mathcal{F}^{12}_{2\bar{3}}|^2 )
= J_{1\bar{1}} ( \mathcal{F}^{12'}_{3\bar{3}} 
\mathcal{F}^{12'}_{2\bar{2}}-|\mathcal{F}^{12'}_{2\bar{3}}|^2 ). 
\label{kahler-9}
\ee

The fluxes along different
branes are constrained in order to satisfy the supersymmetry 
conditions (\ref{kahler-1})-(\ref{kahler-9}). As we are interested
in K\"ahler moduli solutions with vanishing off-diagonal components
$J_{i\bar{j}} = 0$, and using the positivity of the volume form 
$J\wedge J\wedge J$, the above fluxes are restricted as:
\bea
   J_{2\bar{2}}\mathcal{F}^1_{1\bar{1}}\mathcal{F}^1_{3\bar{3}} >  
J_{3\bar{3}} 
\mathcal{F}^1_{1\bar{2}}\mathcal{F}^1_{2\bar{1}},\;\;
   J_{3\bar{3}} \mathcal{F}^2_{2\bar{2}}\mathcal{F}^2_{1\bar{1}} > 
J_{1\bar{1}} 
\mathcal{F}^2_{2\bar{3}}\mathcal{F}^2_{3\bar{2}},\cr
\cr
   J_{1\bar{1}} \mathcal{F}^3_{3\bar{3}}\mathcal{F}^3_{2\bar{2}} > 
J_{2\bar{2}} 
\mathcal{F}^3_{3\bar{1}}\mathcal{F}^3_{1\bar{3}},\;\;
   J_{2\bar{2}} \mathcal{F}^4_{1\bar{1}}\mathcal{F}^4_{3\bar{3}} > 
J_{3\bar{3}}
\mathcal{F}^4_{1\bar{2}}\mathcal{F}^4_{2\bar{1}}, 
\cr
\cr
   J_{3\bar{3}} \mathcal{F}^5_{2\bar{2}}\mathcal{F}^5_{1\bar{1}} > 
J_{1\bar{1}}
\mathcal{F}^5_{2\bar{3}}\mathcal{F}^5_{3\bar{2}},\;\;
   J_{1\bar{1}} \mathcal{F}^6_{3\bar{3}}\mathcal{F}^6_{2\bar{2}} > 
J_{2\bar{2}}
\mathcal{F}^6_{3\bar{1}}\mathcal{F}^6_{1\bar{3}}.
\label{susy-flux1}
\eea
Moreover, for branes 10-12 and $10'$-$12'$ one has to impose:
\bea
  \mathcal{F}^{10}_{3\bar{3}} 
\mathcal{F}^{10}_{1\bar{1}}-|\mathcal{F}^{10}_{1\bar{3}}|^2> 0,\;\;
  \mathcal{F}^{11}_{1\bar{1}} 
\mathcal{F}^{11}_{2\bar{2}}-|\mathcal{F}^{10}_{1\bar{2}}|^2> 0,\;\;
  \mathcal{F}^{12}_{3\bar{3}} 
\mathcal{F}^{12}_{2\bar{2}}-|\mathcal{F}^{12}_{2\bar{3}}|^2  > 0,\cr
\cr
\mathcal{F}^{10'}_{3\bar{3}} 
\mathcal{F}^{10'}_{1\bar{1}}-|\mathcal{F}^{10'}_{1\bar{3}}|^2> 0,\;\;
  \mathcal{F}^{11'}_{1\bar{1}} 
\mathcal{F}^{11'}_{2\bar{2}}-|\mathcal{F}^{11'}_{1\bar{2}}|^2> 0,\;\;
  \mathcal{F}^{12'}_{3\bar{3}} 
\mathcal{F}^{12'}_{2\bar{2}}-|\mathcal{F}^{12'}_{2\bar{3}}|^2  > 0.
\label{susy-flux7}
\eea

We have therefore given a set of conditions to be used 
for solving the supersymmetry equations  
(\ref{kahler-1})-(\ref{kahler-9}). We postpone the  
discussion on their solutions
for the next subsection and examine now the additional 
constraints imposed on fluxes from the 
positivity requirement (\ref{Jcond}).
For  stacks 1-6, this condition reduces to the form
(\ref{Jcond3}). The diagonal fluxes are then restricted to the domain 
where
\bea
 F^1_{x^3y^3} < 0,\;\;F^2_{x^1y^1} < 0,\;\;F^3_{x^2y^2} < 0,\cr
 F^4_{x^3y^3} < 0,\;\;F^5_{x^1y^1} < 0,\;\;F^6_{x^2y^2} < 0.
\label{positivity-1}
\eea
When combined with conditions (\ref{susy-flux1}), this further implies
that the remaining diagonal fluxes in branes 1-6  have to be 
negative, as well:
\bea
 F^1_{x^1y^1} < 0,\;\;F^2_{x^2y^2} < 0,\;\;F^3_{x^3y^3} < 0,\cr
 F^4_{x^1y^1} < 0,\;\;F^5_{x^2y^2} < 0,\;\;F^6_{x^3y^3} < 0.
\label{positivity-2}
\eea
Finally, since the stacks 10-12 and $10'$-$12'$ satisfy
$\mathcal{F}^a\wedge \mathcal{F}^a\wedge  \mathcal{F}^a = 0$,
they do not contribute to the 3-brane charge, and the 
positivity conditions follow from eq.~(\ref{Jcond2}). Combined 
with eq.~(\ref{susy-flux7}), we  get that the magnetic fluxes for 
these branes
must also be negative:
\bea
F^{10}_{x^3y^3} < 0,\;\;\;F^{10}_{x^1y^1} < 0,\;\;\;
F^{10'}_{x^3y^3} < 0,\;\;\;F^{10'}_{x^1y^1} < 0,\cr
F^{11}_{x^1y^1} < 0,\;\;\;F^{11}_{x^2y^2} < 0,\;\;\;
F^{11'}_{x^1y^1} < 0,\;\;\;F^{11'}_{x^2y^2} < 0,\cr
F^{12}_{x^2y^2} < 0,\;\;\;F^{12}_{x^3y^3} < 0,\;\;\;
F^{12'}_{x^2y^2} < 0,\;\;\;F^{12'}_{x^3y^3} < 0.
\label{positivity-3}
\eea

\subsection{Explicit solutions: fluxes and moduli}
\label{subsection-explicit-flux+moduli}

We now present an explicit solution for the fluxes along all stacks 
of  
branes satisfying the restrictions given in equations 
(\ref{consistency}),
(\ref{ki-positive}), (\ref{susy-flux1}), (\ref{susy-flux7})
and (\ref{positivity-1})-(\ref{positivity-3}). These fluxes are
defined in terms of the first Chern numbers $m_i$'s
and winding numbers $n_i$'s, introduced earlier in 
eq.~(\ref{diraca}), along the various 2-cycles of $T^6/\mathbb{Z}_2$. 
We choose for branes 1-3:
\bea
\begin{pmatrix}(m^1_{x^1y^2}, n^1_{x^1y^2}),\;\;\; (m^1_{x^2y^1}, 
n^1_{x^2y^1}) \cr  (m^1_{x^1y^1}, n^1_{x^1y^1}) \cr
                        (m^1_{x^2y^2}, n^1_{x^2y^2}) \cr
                        (m^1_{x^3y^3}, n^1_{x^3y^3})
\end{pmatrix} = 
\begin{pmatrix}
(m^2_{x^2y^3}, n^2_{x^2y^3}),\;\;\; (m^2_{x^3y^2}, n^2_{x^3y^2}) \cr  
       (m^2_{x^2y^2}, n^2_{x^2y^2}) \cr
                        (m^2_{x^3y^3}, n^2_{x^3y^3}) \cr
                        (m^2_{x^1y^1}, n^2_{x^1y^1})
\end{pmatrix} = 
\cr
\begin{pmatrix}(m^3_{x^3y^1}, n^3_{x^3y^1}),\;\;\; (m^3_{x^1y^3}, 
n^3_{x^1y^3}) \cr  
       (m^3_{x^3y^3}, n^3_{x^3y^3}) \cr
                        (m^3_{x^1y^1}, n^3_{x^1y^1}) \cr
                        (m^3_{x^2y^2}, n^3_{x^2y^2})\end{pmatrix} = 
\begin{pmatrix}(-1, 1),\;\;\;(1, -1)\cr
         (2, -1) \cr
         (0, -1) \cr
         (1, -1) \end{pmatrix},
\label{soln.-1}
\eea
Similarly, for branes 4-6 we choose:
\bea
\begin{pmatrix}(m^4_{x^1x^2}, n^4_{x^1x^2}),\;\;\; (m^4_{y^1y^2}, 
n^4_{y^1y^2}) \cr  
        (m^4_{x^1y^1}, n^4_{x^1y^1}) \cr
                        (m^4_{x^2y^2}, n^4_{x^2y^2}) \cr
                        (m^4_{x^3y^3}, n^4_{x^3y^3})\end{pmatrix} = 
\begin{pmatrix}(m^5_{x^2x^3}, n^5_{x^2x^3}),\;\;\; (m^5_{y^2y^3}, 
n^5_{y^2y^3}) \cr  
      (m^5_{x^2y^2}, n^5_{x^2y^2}) \cr
                        (m^5_{x^3y^3}, n^5_{x^3y^3}) \cr
                        (m^5_{x^1y^1}, n^5_{x^1y^1})
                    \end{pmatrix} = \cr
\begin{pmatrix}(m^6_{x^3x^1}, n^6_{x^3x^1}),\;\;\; (m^6_{y^3y^1}, 
n^6_{y^3y^1}) \cr     (m^6_{x^3y^3}, n^6_{x^3y^3}) \cr
                        (m^6_{x^1y^1}, n^6_{x^1y^1}) \cr
                        (m^6_{x^2y^2}, n^6_{x^2y^2})\end{pmatrix} = 
\begin{pmatrix}(-1, 1),\;\;\;(1, -1)\cr
         (2, -1) \cr
         (0, -1) \cr
         (1, -1) \end{pmatrix}.
\label{soln.-2}
\eea
For branes 10-12, the values of the fluxes are given by:
\bea
\begin{pmatrix} (m^{10}_{x^1y^3}, n^{10}_{x^1y^3}),\;\;\; 
(m^{10}_{x^3y^1}, 
n^{10}_{x^3y^1}) \cr 
          (m^{10}_{x^1y^1}, n^{10}_{x^1y^1}) \cr
          (m^{10}_{x^2y^2}, n^{10}_{x^2y^2}) \cr
          (m^{10}_{x^3y^3}, n^{10}_{x^3y^3})
\end{pmatrix} = 
\begin{pmatrix}(m^{11}_{x^2y^1}, n^{11}_{x^2y^1}),\;\;\; 
(m^{11}_{x^1y^2}, 
n^{11}_{x^1y^2}) \cr 
          (m^{11}_{x^2y^2}, n^{11}_{x^2y^2}) \cr
          (m^{11}_{x^3y^3}, n^{11}_{x^3y^3} )\cr
          (m^{11}_{x^1y^1}, n^{11}_{x^1y^1})
\end{pmatrix} = \cr
\begin{pmatrix}(m^{12}_{x^3y^2}, n^{12}_{x^3y^2}),\;\;\; 
(m^{12}_{x^2y^3}, 
n^{12}_{x^2y^3}) \cr 
(m^{12}_{x^3y^3}, n^{12}_{x^3y^3}) \cr
          (m^{12}_{x^1y^1}, n^{12}_{x^1y^1}) \cr
          (m^{12}_{x^2y^2}, n^{12}_{x^2y^2})
\end{pmatrix} = 
\begin{pmatrix}(1,-1),(-1,1) \cr(-1,1) \cr (0, 1)\cr (-2, 
1)\end{pmatrix}\, .
\label{soln.-3}
\eea
Finally, the flux quanta for the stacks $10'$-$12'$ are the same as for 
the stacks 10-12 and are given by:
\bea
\begin{pmatrix} (m^{10'}_{x^1x^3}, n^{10'}_{x^1x^3}),\;\;\; 
(m^{10'}_{y^1y^3}, 
n^{10'}_{y^1y^3}) \cr 
          (m^{10'}_{x^1y^1}, n^{10'}_{x^1y^1}) \cr
          (m^{10'}_{x^2y^2}, n^{10'}_{x^2y^2}) \cr
          (m^{10'}_{x^3y^3}, n^{10'}_{x^3y^3})
\end{pmatrix} = 
\begin{pmatrix}(m^{11'}_{x^2x^1}, n^{11'}_{x^2x^1}),\;\;\; 
(m^{11'}_{y^1y^1}, 
n^{11}_{y^2y^1}) \cr 
          (m^{11'}_{x^2y^2}, n^{11'}_{x^2y^2}) \cr
          (m^{11'}_{x^3y^3}, n^{11'}_{x^3y^3} )\cr
          (m^{11'}_{x^1y^1}, n^{11'}_{x^1y^1})
\end{pmatrix} = \cr
\begin{pmatrix}(m^{12'}_{x^3x^2}, n^{12'}_{x^3x^2}),\;\;\; 
(m^{12'}_{y^3y^2}, 
n^{12'}_{y^3y^2}) \cr 
(m^{12'}_{x^3y^3}, n^{12}_{x^3y^3}) \cr
          (m^{12'}_{x^1y^1}, n^{12}_{x^1y^1}) \cr
          (m^{12'}_{x^2y^2}, n^{12}_{x^2y^2})
\end{pmatrix} = 
\begin{pmatrix}(1,-1),(-1,1) \cr(-1,1) \cr (0, 1)\cr (-2, 
1)\end{pmatrix}\, .
\label{soln.-3a}
\eea

Using the above values of $m$ and $n$, 
the non-zero fluxes defined in eq.~(\ref{diraca})  and  
used in complex structure moduli stabilization of 
section \ref{model-A}, for branes 1-6 read:
\be
\begin{pmatrix}p^1_{x^1y^2}\cr p^1_{x^2y^1}\cr p^1_{x^1y^1}\cr 
p^1_{x^3y^3}\end{pmatrix} =
\begin{pmatrix}p^2_{x^2y^3}\cr p^2_{x^3y^2}\cr p^2_{x^2y^2}\cr 
p^2_{x^1y^1}\end{pmatrix} =
\begin{pmatrix}p^3_{x^3y^1}\cr p^3_{x^1y^3}\cr p^3_{x^3y^3}\cr 
p^3_{x^2y^2}\end{pmatrix} = 
\begin{pmatrix}-1\cr -1\cr -2\cr -1\end{pmatrix},
\label{soln.-flux1}
\ee
\be
\begin{pmatrix}p^4_{x^1x^2}\cr p^4_{y^1y^2}\cr p^4_{x^1y^1}\cr 
p^4_{x^3y^3}\end{pmatrix} =
\begin{pmatrix}p^5_{x^2x^3}\cr p^5_{y^2y^3}\cr p^5_{x^2y^2}\cr 
p^5_{x^1y^1}\end{pmatrix} =
\begin{pmatrix}p^6_{x^3x^1}\cr p^6_{y^3y^1}\cr p^6_{x^3y^3}\cr 
p^6_{x^2y^2}\end{pmatrix} = 
\begin{pmatrix}-1\cr -1\cr -2\cr -1\end{pmatrix}.
\label{soln.-flux2}
\ee
Similarly, for branes 10-12 and $10'$-$12'$, 
the non-zero fluxes are given by:
\be
\begin{pmatrix}p^{10}_{x^1y^3}\cr p^{10}_{x^3y^1}\cr 
p^{10}_{x^1y^1}\cr 
p^{10}_{x^3y^3}\end{pmatrix} =
\begin{pmatrix}p^{11}_{x^1y^2}\cr p^{11}_{x^2y^1}\cr 
p^{11}_{x^2y^2}\cr 
p^{11}_{x^1y^1}\end{pmatrix} =
\begin{pmatrix}p^{12}_{x^2y^3}\cr p^{12}_{x^3y^2}\cr 
p^{12}_{x^3y^3}\cr 
p^{12}_{x^2y^2}\end{pmatrix} = 
\begin{pmatrix}p^{10'}_{x^1x^3}\cr p^{10'}_{y^1y^3}\cr 
p^{10'}_{x^1y^1}\cr 
p^{10'}_{x^3y^3}\end{pmatrix} =
\begin{pmatrix}p^{11'}_{x^1x^2}\cr p^{11'}_{y^1y^2}\cr 
p^{11'}_{x^2y^2}\cr 
p^{11'}_{x^1y^1}\end{pmatrix} =
\begin{pmatrix}p^{12'}_{x^2x^3}\cr p^{12'}_{y^2y^3}\cr 
p^{12'}_{x^3y^3}\cr 
p^{12'}_{x^2y^2}\end{pmatrix} = 
\begin{pmatrix}-1\cr -1 \cr- 1\cr  -2\end{pmatrix}.
\label{soln.-flux3}
\ee

One can now verify that the above magnetic fluxes
(\ref{soln.-flux1})-(\ref{soln.-flux3}) satisfy 
the conditions (\ref{consistency}) and the parameters
$K_i$ defined in eqs.~(\ref{ratio}) and (\ref{product}) read:
\be
K_1 =\;K_2 = \;K_3 =\;K_4 =\; K_5 =\;K_6 = 1,\;\;
\label{value-ki}
\ee
which obviously solve both eqs.~(\ref{consistency}) and
(\ref{ki-positive}). Moreover, the diagonal components of the 
complex structure moduli are fixed using
eq.~(\ref{diagonal-tau}) to:
\be
        \tau^{11} = \tau^{22} = \tau^{33} = i. 
\label{soln.-diagonal-tau}
\ee
Since all diagonal components of the
fluxes (\ref{soln.-flux1})-(\ref{soln.-flux3})
are negative, they also obviously satisfy the conditions
(\ref{positivity-1})-(\ref{positivity-3}). The conditions 
(\ref{susy-flux1}) and (\ref{susy-flux7})
are also satisfied, as  
will be shown in the following 
subsection~\ref{subection-solve-supersymmetry}.

We have therefore shown that the explicit choice for the fluxes
presented in eqs. (\ref{soln.-flux1})-(\ref{soln.-flux3})
satisfy the consistency requirements imposed earlier. 
Obviously, this choice is not unique.
For instance, it is possible to modify them in a way that the 
products  appearing in the 
supersymmetry conditions (\ref{kahler-1})-(\ref{kahler-9}) involving 
also the K\"ahler 
class moduli remain invariant.
Before ending this section, we also give the matrices 
entering in eqs.~(\ref{complex-matrix1})
and (\ref{complex-matrix2}), for the values of fluxes 
(\ref{soln.-flux1})-(\ref{soln.-flux3}). 
The $3\times 3$ matrix appearing in 
the l.h.s. of eq.~(\ref{complex-matrix1}) reads:
\bea
\begin{pmatrix} 0  &  - 1 & 1  \cr
\cr
1 &  0 & -1  \cr               
\cr
- 1 & 1 & 0 
\end{pmatrix},
\label{matrix-value1}
\eea
while the matrix appearing in eq.~(\ref{complex-matrix2})
\bea
\begin{pmatrix} 1 &  0 & - 1  \cr
\cr
0 &  -1  & 1  \cr               
\cr
- 1 & 1 & 0 
\end{pmatrix}
\label{matrix-value2}
\eea
is singular and implies the relation (\ref{tau-0-2}):
$\tau^{21} = \tau^{32} = \tau^{13}$. Using this equality in the 
r.h.s. of eq.~(\ref{complex-matrix1}) with the result
(\ref{matrix-value1}), one finds: $\tau^{12} = \tau^{23} = \tau^{31}$
and $\tau^{21} = \tau^{32} = \tau^{13} = 0$.
Finally, using brane 4 (or brane 5-6) one obtains
the result (\ref{tau-0}) that all off-diagonal 
components of the complex structure $\tau^{ij}$ are zero.

We will now obtain the values of the K\"ahler class moduli
by solving the supersymmetry conditions. We will 
show that all off-diagonal components vanish, while the 
diagonal ones (in real coordinates) are
$J_{x^iy^i} =  (2\pi)^2 \alpha^{\prime}$ (for $i=1,2,3$), defining 
a meaningful solution of the 
supersymmetry conditions (\ref{kahler-1})-(\ref{kahler-9}).

\subsection{Solving the supersymmetry conditions to 
fix the K\"ahler form}\label{subection-solve-supersymmetry}

Here, we analyze the supersymmetry conditions 
(\ref{kahler-1})-(\ref{kahler-9}),
which consist of nine independent non-linear equations for nine variables.
The reason is that the three equations in the r.h.s. of (\ref{kahler-7})-(\ref{kahler-9}), 
related to the stacks 10-$10'$, 11-$11'$ and $12$-$12'$, are trivially 
satisfied because of our choice of fluxes. 
Even if the system could in principle be solved exactly, we only 
present here the 
solution where the off-diagonal components of the K\"ahler form 
vanish, 
$J_{i\bar{j}} =0$ for $i\neq j$. This solution is consistent with 
eqs.~(\ref{kahler-7})-(\ref{kahler-9}), arising from the brane stacks 
10-12 and $10'$-$12'$
with the choice of fluxes given in (\ref{soln.-flux3}), for a 
restricted K\"ahler 
class moduli space where
\be
 J_{1\bar{1}} = J_{2\bar{2}} = J_{3\bar{3}}.
\label{diagonal-J-equal}
\ee
Moreover, the brane stacks 1-6 restrict further the K\"ahler moduli to
\bea
J_{2\bar{2}}\mathcal{F}^1_{1\bar{1}}\mathcal{F}^1_{3\bar{3}}
                   - 
J_{3\bar{3}}\mathcal{F}^1_{1\bar{2}}\mathcal{F}^1_{2\bar{1}}
= 
J_{3\bar{3}}\mathcal{F}^2_{2\bar{2}}\mathcal{F}^2_{1\bar{1}}
                   - 
J_{1\bar{1}}\mathcal{F}^2_{2\bar{3}}\mathcal{F}^2_{3\bar{2}}
=
J_{1\bar{1}}\mathcal{F}^3_{3\bar{3}}\mathcal{F}^3_{2\bar{2}}
                   - 
J_{2\bar{2}}\mathcal{F}^3_{3\bar{1}}\mathcal{F}^3_{1\bar{3}}
=\cr
\cr
J_{2\bar{2}}\mathcal{F}^4_{1\bar{1}}\mathcal{F}^4_{3\bar{3}}
                   - 
J_{3\bar{3}}\mathcal{F}^4_{1\bar{2}}\mathcal{F}^4_{2\bar{1}}
=
J_{3\bar{3}}\mathcal{F}^5_{2\bar{2}}\mathcal{F}^5_{1\bar{1}}
                   - 
J_{1\bar{1}}\mathcal{F}^5_{2\bar{3}}\mathcal{F}^5_{3\bar{2}}
=
J_{1\bar{1}}\mathcal{F}^6_{3\bar{3}}\mathcal{F}^6_{2\bar{2}}
                   - 
J_{2\bar{2}}\mathcal{F}^6_{3\bar{1}}\mathcal{F}^6_{1\bar{3}}
=\cr
\cr
J_{2\bar{2}} (\mathcal{F}^{10}_{3\bar{3}} 
\mathcal{F}^{10}_{1\bar{1}}- |\mathcal{F}^{10}_{1\bar{3}}|^2)
=
J_{3\bar{3}} (\mathcal{F}^{11}_{1\bar{1}} 
\mathcal{F}^{11}_{2\bar{2}}- |\mathcal{F}^{11}_{1\bar{2}}|^2)
=
J_{1\bar{1}} (\mathcal{F}^{12}_{2\bar{2}} 
\mathcal{F}^{12}_{3\bar{3}}- |\mathcal{F}^{12}_{2\bar{3}}|^2)
= \cr
\cr
J_{2\bar{2}} (\mathcal{F}^{10'}_{3\bar{3}} 
\mathcal{F}^{10'}_{1\bar{1}}- |\mathcal{F}^{10'}_{1\bar{3}}|^2)
=
J_{3\bar{3}} (\mathcal{F}^{11'}_{1\bar{1}} 
\mathcal{F}^{11'}_{2\bar{2}}- |\mathcal{F}^{11'}_{1\bar{2}}|^2)
=
J_{1\bar{1}} (\mathcal{F}^{12'}_{2\bar{2}} 
\mathcal{F}^{12'}_{3\bar{3}}- |\mathcal{F}^{12'}_{2\bar{3}}|^2).
\label{flux-equality}
\eea
Using the choice of fluxes (\ref{soln.-flux1}) -  
(\ref{soln.-flux3}), 
we then get a solution for  the diagonal K\"ahler moduli
\be
J_{1\bar{1}} = J_{2\bar{2}} = J_{3\bar{3}} = 
\frac{ (2\pi)^{2}\alpha'}{2},
\label{diagonal-J=1}
\ee
or in terms of real coordinates:
\be
J_{x^1y^1} = J_{x^2y^2} = J_{x^3y^3} = (2\pi)^{2}\alpha'\, .
\label{diagonal-J=1-real}
\ee
To show that the conditions (\ref{susy-flux1}) and 
(\ref{susy-flux7}) are satisfied, we rewrite the fluxes in the 
complex coordinates 
(\ref{2fbasis}), using (\ref{11part}) and 
eqs.~(\ref{soln.-flux1})-(\ref{soln.-flux3}):
\bea
\mathcal{F}^1_{1\bar{2}} =  \mathcal{F}^1_{2\bar{1}} = 
\mathcal{F}^2_{2\bar{3}} =  \mathcal{F}^2_{3\bar{2}} = 
\mathcal{F}^3_{3\bar{1}} =  \mathcal{F}^3_{1\bar{3}} = -{1\over 
2}4\pi^2\alpha',\cr
\cr
\mathcal{F}^4_{1\bar{2}} =  - \mathcal{F}^4_{2\bar{1}} =
\mathcal{F}^5_{2\bar{3}} =  - \mathcal{F}^5_{3\bar{2}} =
\mathcal{F}^6_{3\bar{1}} =  - \mathcal{F}^6_{1\bar{3}} = 2\pi^2i\alpha'\, ,
\label{complex-flux-values1}
\eea
and
\be
\mathcal{F}^1_{1\bar{1}} = \mathcal{F}^2_{2\bar{2}}
= \mathcal{F}^3_{3\bar{3}} = 
\mathcal{F}^4_{1\bar{1}} = \mathcal{F}^5_{2\bar{2}}
= \mathcal{F}^6_{3\bar{3}} = - 4\pi^2\alpha'
\label{complex-flux-values2}
\ee
\be
\mathcal{F}^1_{3\bar{3}} = \mathcal{F}^2_{1\bar{1}}
= \mathcal{F}^3_{2\bar{2}} = 
\mathcal{F}^4_{3\bar{3}} = \mathcal{F}^5_{1\bar{1}}
= \mathcal{F}^6_{2\bar{2}} =  - 2\pi^2\alpha'\, .
\label{complex-flux-values3}
\ee
It is then easy to see that the conditions (\ref{susy-flux1}) are
satisfied, using the result 
$J_{1\bar{1}}=J_{2\bar{2}}=J_{3\bar{3}}$. 
Similarly, the conditions 
(\ref{susy-flux7}) are  satisfied using the following 
expressions for the magnetic fluxes along the branes 10-12 in complex 
coordinates:
\bea
2\mathcal{F}^{10}_{1\bar{1}} = \mathcal{F}^{10}_{3\bar{3}}
= 2\mathcal{F}^{11}_{2\bar{2}} = 
\mathcal{F}^{11}_{1\bar{1}} = 2\mathcal{F}^{12}_{3\bar{3}}
= \mathcal{F}^{12}_{2\bar{2}} = - 4\pi^2\alpha',\cr
\cr
2\mathcal{F}^{10}_{1\bar{3}} = 2\mathcal{F}^{10}_{3\bar{1}}
= 2\mathcal{F}^{11}_{2\bar{1}} = 
\mathcal{F}^{11}_{1\bar{2}} = 2\mathcal{F}^{12}_{3\bar{2}}
= \mathcal{F}^{12}_{2\bar{3}} = - 4\pi^2\alpha', 
\label{complex-flux-values4}
\eea
and for the branes $10'$-$12'$:
\bea
2\mathcal{F}^{10'}_{1\bar{1}} = \mathcal{F}^{10'}_{3\bar{3}}
= 2\mathcal{F}^{11'}_{2\bar{2}} = 
\mathcal{F}^{11'}_{1\bar{1}} = 2\mathcal{F}^{12'}_{3\bar{3}}
= \mathcal{F}^{12'}_{2\bar{2}} = - 4\pi^2\alpha',\cr
\cr
2\mathcal{F}^{10'}_{1\bar{3}} = 2\mathcal{F}^{10'}_{3\bar{1}}
= 2\mathcal{F}^{11'}_{2\bar{1}} = 
\mathcal{F}^{11'}_{1\bar{2}} = 2\mathcal{F}^{12'}_{3\bar{2}}
= \mathcal{F}^{12'}_{2\bar{3}} =  4\pi^2i\alpha'. 
\label{complex-flux-values4'}
\eea

\subsection{Tadpole cancellations}\label{subsection-tadpole-A}

We now analyze the tadpole cancellation conditions, written in 
equations  (\ref{t7}), (\ref{t3}) and (\ref{tad3ff}) 
for model-A, specified
by the  quantum numbers $(m,n)$ of 
eqs.~(\ref{soln.-1})-(\ref{soln.-3a}). 
We start with the analysis of the
7-brane R-R tadpoles (\ref{t7}).
The expressions for the tadpole contributions $q^a_t$
from the $a$-th brane,  localized at the 
2-cycle $C_t^{(2)}$, 
are given in eq.~(\ref{q7R}). For example,  brane-1
has a potential contribution in the following 2-cycles:
\be
q^{1}_{ [x^1y^2]} =
- n^1_{x^3y^3} n^1_{x^2y^1}
 m^1_{x^1y^2} \,\,\, , \,\,\, q^{1}_{ [x^2y^1]} = 
- n^1_{x^3y^3} n^1_{x^1y^2}
 m^1_{x^2y^1} \,\,\, ,
 \ee
 \be
q^{1}_{ [x^1y^1]} = n^1_{x^2y^2} n^1_{x^3y^3} m^1_{x^1y^1} \, \,, \,\,
q^{1}_{ [x^2y^2]} = n^1_{x^1y^1} n^1_{x^3y^3} 
m^1_{x^2y^2}
\,\, , \,\,
q^{1}_{ [x^3y^3]} = 
n^1_{x^1y^1} n^1_{x^2y^2}
 m^1_{x^3y^3}.
\label{7R-tadpole-1}
\ee
By inserting the values of $m$'s and $n$'s from eq.~(\ref{soln.-1}),
we obtain for brane-1:
\be
q^{1}_{ [x^1y^2]} = 1,\;\;q^{1}_{ [x^2y^1]} = 1,\;\;
q^{1}_{ [x^1y^1]} = 2,\;\;q^{1}_{ [x^2y^2]} = 0,\;\;
q^{1}_{ [x^3y^3]} = 1\, ,
\label{tadpole-no.1}
\ee
and similarly for brane-2:
\be
q^{2}_{ [x^2y^3]} = 1,\;\;q^{2}_{ [x^3y^2]} = 1\;\;
q^{2}_{ [x^2y^2]} = 2 ,\;\;q^{2}_{ [x^3y^3]} = 0,\;\;
q^{2}_{ [x^1y^1]} = 1\, ,
\label{tadpole-no.2}
\ee
and brane-3:
\be
q^{3}_{ [x^3y^1]} = 1,\;\;q^{3}_{ [x^1y^3]} = 1\;\;
q^{3}_{ [x^3y^3]} = 2 ,\;\;q^{3}_{ [x^1y^1]} = 0,\;\;
q^{3}_{ [x^2y^2]} = 1\, .
\label{tadpole-no.3}
\ee

The computation is similar for brane-4 to brane-6
with fluxes given in eqs.~(\ref{soln.-2}). The result for the 7-brane
charges is:
\be
q^{4}_{ [x^1x^2]} = 1,\;\;q^{4}_{ [y^1y^2]} = 1,\;\;
q^{4}_{ [x^1y^1]} = 2 ,\;\;q^{4}_{ [x^2y^2]} = 0,\;\;
q^{4}_{ [x^3y^3]} = 1\, ,
\label{tadpole-no.4}
\ee
\be
q^{5}_{ [x^2x^3]} = 1,\;\;q^{5}_{ [y^2y^3]} = 1.\;\;
q^{5}_{ [x^2y^2]} = 2,\;\;q^{5}_{ [x^3y^3]} = 0,\;\;
q^{5}_{ [x^1y^1]} = 1\, ,
\label{tadpole-no.5}
\ee
\be
q^{6}_{ [x^3x^1]} = 1,\;\;q^{6}_{ [y^3y^1]} = 1,\;\;
q^{6}_{ [x^3y^3]} = 2,\;\;q^{6}_{ [x^1y^1]} = 0,\;\;
q^{6}_{ [x^2y^2]} = 1\, .
\label{tadpole-no.6}
\ee

Assuming that each stack  contains  only one brane $N_a = 1$ $\forall 
a$, 
and adding the above contributions to the 7-brane tadpoles from 
branes 
1-6, we obtain a non-vanishing result for 
the diagonal 2-cycles $[x^1 y^1]$, $[x^2 y^2]$, $[x^3 y^3]$:
\be
 \sum_{a=1}^{6}\, N_a\, q^{a}_{t} =  6  \quad , \quad
t = [x^1 y^1], [x^2 y^2], [x^3 y^3]\, .
\label{total-7tadpole,1-6}
\ee
On the other hand, for each of the twelve off-diagonal 2-cycles: $[x^i y^j]$,$[ x^ix^j]$,$[ y^iy^j]$ for $i\neq j$ we have: 
\be
 \sum_{a=1}^{6}\, N_a\, q^{a}_{t} =  1  \quad , \quad
t = [x^i y^j]\, ,\, [x^ix^j]\, , \, [y^iy^j] \,,\;i\neq j. 
\label{total-7tadpole,1-6-off}
\ee

The 7-brane tadpole contributions for the branes 
10-12 with fluxes and quantum numbers $(m, n)$ given in 
eq.~(\ref{soln.-3})
are also non-vanishing and read:
\bea
q^{10}_{ [x^1y^1]} = n^{10}_{x^2y^2} n^{10}_{x^3y^3}m^{10}_{x^1y^1} = 
-1 \,\,\, , \,\,\,
q^{10}_{ [x^3y^3]} = n^{10}_{x^1y^1} n^{10}_{x^2y^2}m^{10}_{x^3y^3} = 
-2\, ,\cr
q^{10}_{ [x^1y^3]} = - n^{10}_{x^2y^2} n^{10}_{x^3y^1}m^{10}_{x^1y^3} = 
-1 \,\,\, , \,\,\,
q^{10}_{ [x^3y^1]} = - n^{10}_{x^2y^2} n^{10}_{x^1y^3}m^{10}_{x^3y^1} = 
-1\, ,
\label{tadpole-no.7}
\eea
\bea
q^{{11}}_{ [x^2y^2]} = n^{11}_{x^3y^3} n^{11}_{x^1y^1}m^{11}_{x^2y^2} 
= 
-1 \,\,\, , \,\,\,
q^{{11}}_{ [x^1y^1]} = n^{11}_{x^2y^2} n^{11}_{x^3y^3}m^{11}_{x^1y^1} 
= 
-2,\cr
q^{{11}}_{ [x^2y^1]} = n^{11}_{x^3y^3} n^{11}_{x^1y^2}m^{11}_{x^2y^1} 
= 
-1 \,\,\, , \,\,\,
q^{{11}}_{ [x^1y^2]} = n^{11}_{x^2y^2} n^{11}_{x^2y^1}m^{11}_{x^1y^2} 
= 
-1,
\label{tadpole-no.8}
\eea
\bea
q^{12}_{ [x^3y^3]} = n^{12}_{x^1y^1} n^{12}_{x^2y^2}m^{12}_{x^3y^3} = 
-1 \,\,\, ,\,\,\,
q^{12}_{ [x^2y^2]} = n^{12}_{x^3y^3} n^{12}_{x^1y^1}m^{12}_{x^2y^2} = 
-2,
\cr
q^{12}_{ [x^3y^2]} = - n^{12}_{x^1y^1} n^{12}_{x^2y^3}m^{12}_{x^3y^2} = 
-1 \,\,\, ,\,\,\,
q^{12}_{ [x^2y^3]} = - n^{12}_{x^1y^1} n^{12}_{x^3y^2}m^{12}_{x^2y^3} = 
-1.
\label{tadpole-no.9}
\eea

Similarly, the contributions of the branes $10'$-$12'$ read:

\be
q^{10'}_{ [x^1y^1]} =  -1 \,\,\, , \,\,\,
q^{10'}_{ [x^3y^3]} =  -2\, 
q^{10'}_{ [x^1x^3]} =  -1 \,\,\, , \,\,\,
q^{10'}_{ [y^1y^3]} =  -1\, ,
\label{tadpole-no.7'}
\ee
\be
q^{{11'}}_{ [x^2y^2]} = -1 \,\,\, , \,\,\,
q^{{11'}}_{ [x^1y^1]} = -2,\,
q^{{11'}}_{ [x^1x^2] }= -1 \,\,\, , \,\,\,
q^{{11'}}_{ [y^1y^2]} = -1\, ,
\label{tadpole-no.8'}
\ee
\be
q^{12'}_{ [x^3y^3]}   = -1 \,\,\, ,\,\,\,
q^{12'}_{ [x^2y^2]}   = -2,\,
q^{12'}_{ [x^2x^3]}   = -1 \,\,\, ,\,\,\,
q^{12'}_{ [y^2y^3]}   = -1.
\label{tadpole-no.9'}
\ee
Adding the results of eqs.~(\ref{tadpole-no.7})-(\ref{tadpole-no.9'}),
we obtain the (non-vanishing) 
contributions of branes 10-12 and $10'$-$12'$ to the 7-brane tadpoles:
\be
 q_{7,R}^t = \sum_{a=10}^{12}\, N_a\, q^{a}_{t}+\sum_{a=10'}^{12'}\, 
N_a\, q^{a}_{t} =-6  \quad , \quad
t = [x^1 y^1], [x^2 y^2], [x^3 y^3]\, .
\label{total-7tadpole,7-9}
\ee
For off-diagonal 2-cycles 
one now has: 
\be
 \sum_{a=10}^{12}\, N_a\, q^{a}_{t} 
+ \sum_{a=10'}^{12'}\, N_a\, q^{a}_{t}
=  -1  \quad , \quad
t = [x^i y^j] \,, \,[ x^ix^j]\, , \,[ y^iy^j]\, , \;i\neq j. 
\label{total-7tadpole,7-9-off}
\ee
{}From eqs.~(\ref{total-7tadpole,1-6}), (\ref{total-7tadpole,1-6-off})
and (\ref{total-7tadpole,7-9}), (\ref{total-7tadpole,7-9-off}), 
we conclude that the total 7-brane 
R-R tadpoles vanish for all 2-cycles, when the contributions of
all the branes are added.

Let us now discuss the 3-brane tadpole cancellation. 
It can be directly verified, using  (\ref{t3}), the brane 
multiplicities $N_a=1$ for each of the twelve stacks, and the
quantum numbers $m$ specified  in 
eqs.~(\ref{soln.-1})-(\ref{soln.-3a}) that each of the branes 1-6 
contributes 
$q^a_{3,R}=1$ ($a = 1,..,6$) to the 3-brane 
tadpole, whereas for branes 10-12 and $10'$-$12'$ one obtains a vanishing 
contribution $q^a_{3,R}=0$
($a =10$-$12$, $10'$-$12'$). The total 3-brane tadpole contribution 
is therefore equal to $\sum_{a=1}^{12} q^a_{3, R} =6$. One 
possibility to cancel the 
3-brane tadpole, namely to satisfy eq. (\ref{t3}), is to either
take multiple copies of various branes 1-6, or/and add 
ordinary $D3$ branes to the system. On the other hand, one can also 
cancel the 3-brane tadpoles by turning on closed string 3-form fluxes,
as discussed in section \ref{sec:3ff}. In this way, one also has the  
advantage that the remaining closed string moduli,
corresponding to the axion and dilaton, can be stabilized as well, 
specifying the string coupling uniquely.

\subsection{Stabilization of the axion-dilaton 
moduli}\label{sec:A-3-form}

As shown above, the model presented in section 
\ref{subsection-explicit-flux+moduli} 
is   a consistent supersymmetric four dimensional perturbative vacuum 
with all 
closed
string moduli but the dilaton fixed. In particular, the metric moduli,
represented by the complex structure and the K\"ahler class are 
stabilized at
the values
\be
J_{i\bar{j}} = \frac{4\pi^2\alpha'}{2}\delta_{i\bar{j}} \quad , \quad
\tau^{ij} = i \delta^{ij}.
\label{ld:kc}
\ee
We want to address now the question of the dilaton stabilization
\cite{Frey:2002hf} by turning on  R-R and NS-NS
3-form fluxes, as explained in the section \ref{sec:3ff}.
Consider  the case where we switch on the following quanta
\be
h_1^{12} \quad , \quad h_1^{21}\quad , \quad  h_2^{12} \quad , \quad 
h_2^{21}\, ; 
\ee
\be
f_1^{12} \quad , \quad f_1^{21}\quad , \quad  f_2^{12} \quad , \quad 
f_2^{21}\, . 
\ee
Since the complex structure has been fixed to the purely imaginary 
diagonal
form (\ref{ld:kc}),  the supersymmetry conditions (\ref{3ffcond1}) 
and 
(\ref{3ffcond2}) are trivially satisfied whereas eq.~(\ref{3ffcond3}) 
fixes the dilaton in terms of the complex structure element 
$\tau_{33} = i$:
\be
f_{2}^{12}-\phi h_{2}^{12} = \tau_{33} (f_{1}^{21}-\phi h_{1}^{21}) 
\quad , 
\quad f_{2}^{21}-\phi h_{2}^{21} =  \tau_{33}(f_{1}^{12}-\phi 
h_{1}^{12}).
\label{dilcond1}
\ee
The two equations can not be independent and 
give rise to a constraint on the allowed fluxes. Actually, these 
conditions 
are  equivalent to the requirement that the $G_{(3)}$ flux must be of 
the type 
$(2,1)$. Indeed, in the complex coordinates 
(\ref{complex_structure}), the only 
non-vanishing
components of $G_{(2,1)}$ are  $G_{1\bar{1}\bar{3}}$ and 
$G_{2\bar{2}\bar{3}}$ :
\be
-2iG_{(2,1)} =( f_{1}^{21}-\phi h_{1}^{21} ) dz_{1}\wedge 
d\bar{z}_{\bar{1}}\wedge dz_{3}
-(f_{1}^{12}-\phi h_{1}^{12}) dz_{2}\wedge 
d\bar{z}_{\bar{2}}\wedge dz_{3}.
\ee
Furthermore, for the values (\ref{ld:kc}) of the K\"ahler form , 
the primitivity condition $G_{(3)}\wedge 
J = 0$ restricts further the fluxes to 
\be
(f_{1}^{12}-\phi h_{1}^{12})J_{1\bar{1}} - (f_{1}^{21}-\phi 
h_{1}^{21})J_{2\bar{2}} = 0, \quad {\rm or \,\, equivalently} \quad 
(f_{1}^{12}-\phi 
h_{1}^{12}) = (f_{1}^{21}-\phi 
h_{1}^{21}).
\label{dilrestr1}
\ee
A quick computation shows that  under the restriction on the fluxes 
coming from 
eqs.~(\ref{dilcond1}) and from the primitivity condition 
(\ref{dilrestr1}), the string coupling is 
given by:
\be
\frac{1}{g_{s}} = {\rm Im }\tau_{33}\frac{f_{2}^{21}h_{1}^{12} - 
f_{1}^{12}h_{2}^{21}}{(h_{1}^{12}{\rm Im }\tau_{33})^{2} + 
(h_{2}^{21})^{2}},
\label{string-coupling}
\ee
if we assume that the complex structure modulus $\tau_{33}$ has no 
real part, as we found in model-A. However, we keep track of the 
dependence of $g_s$ on the complex 
structure in order to examine (in the next section) 
the possible stabilization of the string coupling at small values. 

To simplify the discussion, let us further reduce the number of flux 
components to the case where 
only four of them are different than zero:
\be
  f_1^{12},\;\;f_1^{21},\;\;h_2^{12},\;\;h_2^{21} \neq 0.
\label{dil:flux2}
\ee
Notice that this restriction is only 
possible in the
absence of a B-field, as explained in section \ref{sec:NSNSB}.
Indeed, in the presence of a $B$-field $(b=1/2)$, the flux components 
$h_1^{12}$ and
$h_1^{21}$ have to be odd and can therefore not be set to zero. With 
the
reduced number of non-vanishing elements (\ref{dil:flux2}), the 
string coupling 
(\ref{string-coupling}) is then fixed to the
value 
\be
\frac{1}{g_s} = -{\rm Im }\tau_{33}\frac{f_{1}^{12}}{h_2^{21}}, 
\label{string-coupling2}
\ee
whereas the flux restrictions (\ref{dilcond1}) and
(\ref{dilrestr1}) lead to:
\be
f_1^{12} = f_1^{21} \quad , \quad h_2^{21} = h_2^{12}.
\label{dilrestr2}
\ee
In order to compute the value of the dilaton, we first have to 
analyze the
3-form contribution to the 3-brane tadpole (\ref{N3pos}). From the 
symplectic
structure (\ref{symp_st}) and the restriction
(\ref{dilrestr2}), the 3-form fluxes (\ref{dil:flux2}) induce a 
3-brane charge
\be
N_3 = -\frac{1}{2}(h_2^{12}f_1^{12} + h_2^{12}f_1^{12}) = 
-h_2^{12}f_1^{12}, 
\ee 
and using the results of 
section \ref{subsection-tadpole-A}, the tadpole condition 
(\ref{tad3ff}) reads: 
\be
-h_2^{12}f_1^{12} + N_{D3}= 10.
\label{tad6final}
\ee
As the flux quanta (\ref{dil:flux2}) have to be even in the absence 
of 
a B-field, the minimal value we can get for the string coupling is 
given by
$f_{1}^{12}= -4$ and $h_2^{12}= 2$, which (for $\tau_{33} = i$) 
corresponds to $g_s = 1/2$.

In fact, a smaller value of $g_s$ can be obtained in the presence of a
non-trivial NS-NS $B$ field. Consider for instance the case where a
$B_{x_3y_3} \neq 0$ is introduced in the third torus. As explained in 
section \ref{sec:NSNSB}, the presence of $b=1/2$ induces a
different quantization of the flux quanta (\ref{oddquanta}), which 
can now be odd integers. Moreover, the 3-brane tadpole condition 
(\ref{tad3ff}) is
modified to (\ref{t3NSNSB}).  Since the first Chern number 
$m_{x_3y_3}$ 
is shifted by $b=1/2$, its minimal value is $m_{x_3y_3} = 1/2$. 
It is therefore in principle possible to find a
model similar to Model-A, where the six stacks of branes contribute 
half of
the previous 3-brane charge ${\tilde q}_{3,R} = 3$ instead of $6$. 
The tadpole condition  (\ref{t3NSNSB}) then reads
\be
N_{3} + N_{D3} = 5 \label{N3NSNSB}.
\ee
Assuming that the only non-vanishing components of the 3-form fluxes 
are 
still the ones of eq.~(\ref{dil:flux2}), the
tadpole condition (\ref{N3NSNSB}) becomes
\be
-h_2^{12}f_1^{12} + N_{D3} = 5.
\ee
Since the quantum of the R-R flux $f_1^{12}$ has to be even, the 
minimal value for the string
coupling $g_s$  is given by the choice of fluxes $h_2^{12} = 1$ and 
$f_1^{12} = -4$, 
with $N_{D3}=1$. It follows from eq.~(\ref{string-coupling2}), that 
the
string coupling is then stabilized to the value $g_s=1/4$.

\subsection{Another possibility: Model-A$'$}\label{subsec:nongeom}

In this section, we present another supersymmetric solution with total 
3-brane tadpole contribution  $q_{3,R}=6$ to the 3-brane charge, 
vanishing 7-brane charges and complex structure and K\"ahler moduli 
fixed at the same values as before:
\be
\tau_{ij} = i \delta_{ij} \,\,\, , \,\,\, 
J_{i\bar{j}}={(2\pi)^2\alpha'\over 2}\delta_{i\bar{j}}.
\label{nongeom:t+j}
\ee
However, instead of introducing 12 stacks of branes, we only introduce 
nine, namely the stacks 1 to 9 
given in eqns. (\ref{no.1})-(\ref{no.6}) and (\ref{no.7})-(\ref{no.9}),
relaxing the ``geometric" constraint (\ref{intersection2}) but keeping all
intersection numbers to integer values. 

As non-trivial flux configurations,
we choose for branes 1-3:
\bea
\begin{pmatrix}(m^1_{x^1y^2}, n^1_{x^1y^2}),\;\;\; (m^1_{x^2y^1}, 
n^1_{x^2y^1}) \cr  
        (m^1_{x^2y^3}, n^1_{x^2y^3}),\;\;\;  (m^1_{x^3y^2}, 
n^1_{x^3y^2}) \cr
        (m^1_{x^1y^3}, n^1_{x^1y^3}),\;\;\; (m^1_{x^3y^1}, 
n^1_{x^3y^1})  \cr  
                        (m^1_{x^1y^1}, n^1_{x^1y^1}) \cr
                        (m^1_{x^2y^2}, n^1_{x^2y^2}) \cr
                        (m^1_{x^3y^3}, n^1_{x^3y^3})\end{pmatrix} = 
\begin{pmatrix}(m^2_{x^2y^3}, n^2_{x^2y^3}),\;\;\; (m^2_{x^3y^2}, 
n^2_{x^3y^2}) \cr  
        (m^2_{x^3y^1}, n^2_{x^3y^1}),\;\;\;  (m^2_{x^1y^3}, 
n^2_{x^1y^3}) \cr
        (m^2_{x^2y^1}, n^2_{x^2y^1}),\;\;\; (m^2_{x^1y^2}, 
n^2_{x^1y^2})  \cr  
                        (m^2_{x^2y^2}, n^2_{x^2y^2}) \cr
                        (m^2_{x^3y^3}, n^2_{x^3y^3}) \cr
                        (m^2_{x^1y^1}, n^2_{x^1y^1})\end{pmatrix} = 
\cr
\begin{pmatrix}(m^3_{x^3y^1}, n^3_{x^3y^1}),\;\;\; (m^3_{x^1y^3}, 
n^3_{x^1y^3}) \cr  
        (m^3_{x^1y^2}, n^3_{x^1y^2}),\;\;\;  (m^3_{x^2y^1}, 
n^3_{x^2y^1}) \cr
        (m^3_{x^3y^2}, n^3_{x^3y^2}),\;\;\; (m^3_{x^2y^3}, 
n^3_{x^2y^3})  \cr  
                        (m^3_{x^3y^3}, n^3_{x^3y^3}) \cr
                        (m^3_{x^1y^1}, n^3_{x^1y^1}) \cr
                        (m^3_{x^2y^2}, n^3_{x^2y^2})\end{pmatrix} = 
\begin{pmatrix}(-1, 1),\;\;\;(1, -1)\cr
          (0,1),\;\;\;(0, -1) \cr
         (0, 1),\;\;\;(0, 1) \cr
         (2, -1) \cr
         (0, l) \cr
         (1, -1) \end{pmatrix},
\label{soln.-1'}
\eea
where the integer $l$ in the last column specifies the numerical 
value of the 
windings along the indicated 2-cycles. For the time being, $l$ is 
left arbitrary. As we will see later on, this parameter does not 
affect any of the previous discussions on moduli 
stabilization,
as the flux along this particular 2-cycle is zero. 
Similarly, for branes 4-6 we choose:
\bea
\begin{pmatrix}(m^4_{x^1x^2}, n^4_{x^1x^2}),\;\;\; (m^4_{y^1y^2}, 
n^4_{y^1y^2}) \cr  
        (m^4_{x^2x^3}, n^4_{x^2x^3}),\;\;\;  (m^4_{y^2y^3}, 
n^4_{y^2y^3}) \cr
        (m^4_{x^3x^1}, n^4_{x^3x^1}),\;\;\; (m^4_{y^3y^1}, 
n^4_{y^3y^1})  \cr
        (m^4_{x^3y^1}, n^4_{x^3y^1}),\;\;\; (m^4_{x^2y^3}, 
n^4_{x^2y^3})  \cr
                        (m^4_{x^1y^1}, n^4_{x^1y^1}) \cr
                        (m^4_{x^2y^2}, n^4_{x^2y^2}) \cr
                        (m^4_{x^3y^3}, n^4_{x^3y^3})\end{pmatrix} = 
\begin{pmatrix}(m^5_{x^2x^3}, n^5_{x^2x^3}),\;\;\; (m^5_{y^2y^3}, 
n^5_{y^2y^3}) \cr  
        (m^5_{x^3x^1}, n^5_{x^3x^1}),\;\;\;  (m^5_{y^3y^1}, 
n^5_{y^3y^1}) \cr
        (m^5_{x^1x^2}, n^5_{x^1x^2}),\;\;\; (m^5_{y^1y^2}, 
n^5_{y^1y^2})  \cr
        (m^5_{x^1y^2}, n^5_{x^1y^2}),\;\;\; (m^5_{x^3y^1}, 
n^5_{x^3y^1})  \cr
                        (m^5_{x^2y^2}, n^5_{x^2y^2}) \cr
                        (m^5_{x^3y^3}, n^5_{x^3y^3}) \cr
                        (m^5_{x^1y^1}, n^5_{x^1y^1})
                    \end{pmatrix} = \cr
\begin{pmatrix}(m^6_{x^3x^1}, n^6_{x^3x^1}),\;\;\; (m^6_{y^3y^1}, 
n^6_{y^3y^1}) \cr  
        (m^6_{x^1x^2}, n^6_{x^1x^2}),\;\;\;  (m^6_{y^1y^2}, 
n^6_{y^1y^2}) \cr
        (m^6_{x^2x^3}, n^6_{x^2x^3}),\;\;\; (m^6_{y^2y^3}, 
n^6_{y^2y^3})  \cr  
        (m^6_{x^2y^3}, n^6_{x^2y^3}),\;\;\; (m^6_{x^1y^2}, 
n^6_{x^1y^2})  \cr
                        (m^6_{x^3y^3}, n^6_{x^3y^3}) \cr
                        (m^6_{x^1y^1}, n^6_{x^1y^1}) \cr
                        (m^6_{x^2y^2}, n^6_{x^2y^2})\end{pmatrix} = 
\begin{pmatrix}(-1, 1),\;\;\;(1, -1)\cr
          (0,1),\;\;\;(0, -1) \cr
         (0, 1),\;\;\;(0, 1) \cr
          (0, 1),\;\;\;(0, 1)\cr
         (2, -1) \cr
         (0, l) \cr
         (1, -1) \end{pmatrix}.
\label{soln.-2'}
\eea
Finally, for branes 7-9, the values of the fluxes are given by:
\bea
\begin{pmatrix}(m^7_{x^1y^1}, n^7_{x^1y^1}) \cr
          (m^7_{x^2y^2}, n^7_{x^2y^2}) \cr
          (m^7_{x^3y^3}, n^7_{x^3y^3})\end{pmatrix} = 
\begin{pmatrix}(m^8_{x^2y^2}, n^8_{x^2y^2}) \cr
          (m^8_{x^3y^3}, n^8_{x^3y^3} )\cr
          (m^8_{x^1y^1}, n^8_{x^1y^1})\end{pmatrix} = 
\begin{pmatrix}(m^9_{x^3y^3}, n^9_{x^3y^3}) \cr
          (m^9_{x^1y^1}, n^9_{x^1y^1}) \cr
          (m^9_{x^2y^2}, n^9_{x^2y^2})\end{pmatrix} = 
        \begin{pmatrix}(-1,1) \cr (0, 3(-l+1))\cr (-1, 
1))\end{pmatrix}\, .
\label{soln.-3'}
\eea

It is easy to see that this configuration of fluxes satisfy the consistency 
conditions imposing the absence of antibranes (\ref{no_anti}) 
or (\ref{no_anti2}), the positivity condition  (\ref{Jcond}) and the 
supersymmetry conditions (\ref{kc1}) and (\ref{csc1}) for the values 
of the moduli given in (\ref{nongeom:t+j}). However, the condition 
(\ref{intersection2}) is obviouly not satisfied for generic values of 
the parameter $l$; it is satisfied only for the values $l=\pm 1$.
Despite this fact, the intersection numbers 
(\ref{intersection}) are integers for any pair of stacks presented in 
(\ref{soln.-1'})-(\ref{soln.-3'}). Furthermore, all tadpole conditions 
are satisfied:
\begin{itemize}
\item The first six stacks give rise to a 3-brane charge $q_{3,R}=6$ 
for the simple case in which each stack is composed of a single brane. 
\item The 7-brane charges induced in the diagonal directions $t = 
[x^1y^1],[x^2y^2],[x^3y^3]$ from the 6 first stacks are canceled by the 
choice of fluxes in the last three stacks. 
\item 7-brane charges along the off-diagonal directions,  $t = 
[x^iy^j],[x^ix^j],[y^iy^j]$ where $i\neq j$, can be in principle induced only by 
the stacks 1-6, since branes 7-9 have only diagonal fluxes. However, we 
have chosen the winding numbers in 
(\ref{soln.-1'}) and (\ref{soln.-2'}) in such a way, so that the 
effective winding around the 4-cycle perpendicular to each 
2-cycle $C_t^{(2)}$ vanishes. Thus, all off-diagonal contributions
vanish for each brane separately. 
\end{itemize} 

This model therefore satisfies all consistency conditions listed in 
section \ref{sec:tadpoles}. Its special feature is the 
presence of an additional parameter $l$ that represents the winding number of some 
2-cycles where the branes have vanishing first Chern number. As a 
consequence, it does affect neither the magnetic fluxes $F_r = {m_r 
\over n_r}$ nor the supersymmetry conditions (\ref{kc1})-(\ref{csc1}) 
and therefore does not change the values of the fixed moduli. 
However, since the tadpoles $q^t$ and the overall winding number 
$W_a$  of the brane stacks are sensitive to $l$, the different 
vacua will have different couplings and spectra. Thus, the presence of this 
parameter implies the existence of an infinite family 
of vacua with identical values for the geometrical moduli but with 
different couplings and spectra.  It is therefore important to check further
the consistency of this model by computing for instance its partition function.

\section{Large dimensions}\label{sec:led}

Here, we examine  the possibility to stabilize 
the transverse to the $D3$ branes volume modulus at large values. 
In the T-dual case presented in \cite{AM}, for a supersymmetric 
vacuum compatible with the 
presence of $O9$-planes, it is possible to obtain for instance two 
large radii longitudinal to the magnetized $D9$ branes. This was 
achieved by an 
appropriate rescaling of the magnetic fluxes $m_r$, which is 
compatible with all tadpole cancellation conditions. On the other 
hand, the winding numbers $n_r$ can not be rescaled, because they are 
constrained by  the 9-brane tadpole condition. Similarly, by a 
uniform rescaling of all magnetic
fluxes, one could obtain a family of solutions with all six radii 
large.

The situation in our case is similar. The vacuum presented in the 
section \ref{subsection-explicit-flux+moduli} corresponds to the case 
of three orthogonal tori $T^2 \times T^2 \times T^2$
with radii $R_i^1$ and $R_i^2$. The K\"ahler form and complex 
structure
(\ref{ld:kc}) correspond  to the areas  and  ratios :
\be
J_{x_iy_i} = 4\pi^2 R_i^1R_i^2 \quad , \quad \tau_{ii} = 
i\frac{R^2_i}{R^1_i}
\quad , \quad i=1,2,3.
\label{led:j+t}
\ee
Unlike the T-dual case, now  the 3-brane tadpole condition
(\ref{t3}) restricts strongly the possible rescaling of the first 
Chern
numbers $m_r$, but  it does not constrain the winding numbers. 
There exists therefore a set of different families of an infinite but
discrete number of vacua, 
starting for instance from those found in the previous section:

\begin{itemize}
\item
All radii are rescaled uniformly at values lower than the string 
length $\sqrt{\alpha'}$. 
Thus, K\"ahler moduli are rescaled whereas the complex structure 
remains at the original value:
\be
\hat{J}_r = (2\pi)^2 \Lambda^{-1} \alpha' \,\,\, , \,\,\, \tau_{ij} = 
i\delta_{ij}\, .
\ee
This is achieved be a  rescaling of all winding 
numbers $\hat{n}^a_r = \Lambda n^a_r$, resulting into a  decrease of 
all magnetic 
fluxes
$\hat{F}_r^a = \Lambda^{-1} F_r^a$. Indeed, the supersymmetry 
condition 
(\ref{kc1a})
is then satisfied by rescaled K\"ahler moduli $\hat{J}_r =
\Lambda^{-1} J_r$, $\forall r$. On the other hand, the complex 
structure 
moduli in eq.~(\ref{diagonal-tau}) are given by  ratios of fluxes. 
Therefore,
a general rescaling of the latter does not affect the complex 
structure. 
As a result, the radii of the
different tori $T^2_i$  remain equal even after the rescaling:
$\hat{R}^2_i = \hat{R}^1_i$.  This rescaling is also compatible 
with   the 
7-brane tadpoles. In fact, in 
the setup presented in  section \ref{subsection-tadpole-A}, the  
7-brane
charges induced by the stacks 1 to 6 are cancelled by the 
contributions of the
stacks 7 to 9. As all 7-brane charges (\ref{q7R}) are quadratic in 
the winding
numbers, the tadpole conditions (\ref{t7}) are  left invariant  after 
the rescaling. It is therefore possible to obtain
arbitrary small radii $\hat{R}_i^{2} = {1\over\sqrt{\Lambda}}R_i^2$ 
by a
general rescaling of all winding numbers.

It follows that from the explicit example of a supersymmetric 
vacuum
with fixed moduli (\ref{ld:kc}), there exists  an
infinity of discrete supersymmetric vacua with the same complex 
structure
$\tau_{ij}=i \delta_{ij}$ and
arbitrary small volume moduli $\hat{J}_{x_iy_i}$, $i=1,2,3$. 
It is easy to see that in the T-dual version, this corresponds 
actually to 
large ``longitudinal" dimensions, along the world volume of the 
branes.

\item  A single radius is smaller than the string length 
$\sqrt{\alpha'}$, say $R_3^1$,
whereas the others remain of order of the string length.
This corresponds to the case where the K\"ahler class moduli 
$J_{x_1y_1}$ and $J_{x_2y_2}$ remain fixed, as well as $\tau_{11}$ 
and $\tau_{22}$, whereas the area $J_{x_3y_3}$ of the third $T^2$ is 
small and its radii ratio $\tau_{33}$ is big: 
\be
\tau_{11} = \tau_{22} = i \,\, ,\,\, J_{x_1y_1} = J_{x_2y_2} = 
(2\pi)^2\alpha'\,\,\,\, ; \,\,\,\,\hat{\tau}_{33} = i\Lambda \,\, , 
\hat{J}_{x_3y_3} = (2\pi)^2\Lambda^{-1}\alpha' \, .
\label{led:tau}
\ee
This can be achieved  by a rescaling of the windings of model-A which 
involves the direction $x_3$,\footnote{Note that the coordinate $x_3$ 
in (\ref{led:j+t}) has periodicity $x_3\equiv x_3 +2\pi R_3^1$.} 
namely $\hat{n}^a_{x_3y_i} = \Lambda n^a_{x_3y_i}$ and  
$\hat{n}^a_{x_3x_i} = \Lambda n^a_{x_3x_i}$, for $i=1,2,3$ and for 
all stacks of branes $a = 1,\dots,8$. Indeed, from  
eqs.~(\ref{ratio}) and (\ref{product}), we notice that the complex 
structure moduli $\tau_{11}$ and $\tau_{22}$ are not rescaled, in 
contrast to $\tau_{33}$ which gets rescaled as $\hat{\tau}_{33} = 
\Lambda\tau_{33}$. Furthermore, the solutions to the supersymmetry 
conditions (\ref{kahler-1})-(\ref{kahler-9}) remain valid for a 
rescaled area $\hat{J}_{x_3y_3} = \Lambda^{-1}J_{x_3y_3} $. By a 
similar argument as in the previous case, it can be finally checked 
that even with the rescaled winding numbers, the tadpole cancellation 
conditions (\ref{t3}) and (\ref{t7}) are still satisfied.

This  family of discrete vacua provides a new interesting feature: It
allows the rescaling of the string coupling (\ref{string-coupling2}) 
$g_s
= -\Lambda^{-1}\frac{h_2^{21}}{f_1^{12}}$ without spoiling the tadpole
condition (\ref{tad6final}). This does not come from a rescaling of 
the
3-form quanta $h_2^{21}$ and $f_1^{12}$ and therefore its tadpole
contribution to (\ref{tad6final}) remains invariant (of order unity). 
Note
however that upon T-duality where the small dimension becomes large 
longitudinal, the string coupling becomes again of order one.

\item  Two (or three) of the complex structure moduli are big and two 
(or three) of the $T^2$'s areas become smaller than the string scale. 
For instance, 
\be
\tau_{11}=i \, , \,\,J_{x_1y_1} = (2\pi)^2\alpha' \, ; \, \tau_{22} = 
\tau_{33} = i\Lambda \,\, , \, \hat{J}_{x_2y_2}=\hat{J}_{x_3y_3} = 
(2\pi)^2\Lambda^{-1}\alpha'\, .
\ee
In this example, the radii $R_3^1$ and $R_2^1$ are fixed to a value 
smaller than the string length, keeping the other ones of order 
$\sqrt{\alpha'}$. This can be achieved by the rescaling of all 
winding numbers involving the directions $x_3$ and $x_2$, namely 
\be
\hat{n}^a_{x_2x_3}= \Lambda^2 n^a_{x_2x_3}\, \, , \,\, 
\hat{n}^a_{x_jy_i}=\Lambda n^a_{x_jy_i} \, \, , 
\,\,\hat{n}^a_{x_jx_1}=\Lambda n^a_{x_jx_1}\,\, {\rm for }\,\, 
i=1,2,3 , \, j=2,3\, .
\ee 

\item The areas can be fixed at small values while keeping the radii 
ratios fixed. For instance, we can rescale one area, say of the last 
$T^2$:
\be
\tau_{ij} = i\delta_{ij} \,\,\, , \,\, J_{x_1y_1} = J_{x_2y_2} = 
(2\pi)^2\alpha' \,\, , \,\, J_{x_3y_3} = (2\pi)^2 
\Lambda^{-2}\alpha'\, .
\ee
Here, the radii $R_3^1$ and $R_3^2$ are increased by the rescaling of 
all winding numbers which involves the directions $x_3$ and $y_3$, 
namely
 \be
\hat{n}^a_{x_3y_3}= \Lambda^2 n^a_{x_3y_3}\, \, , \,\, 
\hat{n}^a_{x_iy_3}=\Lambda n^a_{x_iy_3} \, \, , 
\,\,\hat{n}^a_{x_3y_i}=\Lambda n^a_{x_3y_i}\,\, , \,\, 
\hat{n}^a_{x_3x_i}=\Lambda n^a_{x_3x_i} \,\, , \,\, 
\hat{n}^a_{y_3y_i}=\Lambda n^a_{y_3y_i} ,
\ee 
for $i=1,2$.  The same method can be used in order to fix more than 
two areas at values much lower than the string scale $\alpha'$. 
\end{itemize}

\section{Model-B with $q_3 = 12$ }\label{sec:Model-B}

We now present another consistent model for the stabilization
of K\"ahler and complex structure moduli using open and closed 
string fluxes.
In this example, as seen by comparing eqs.~(\ref{no.1})-(\ref{no.6})
with (\ref{no.1'})-(\ref{no.6'}),
certain components of fluxes
(of the type $p^a_{x^iy^j}$, $p^a_{x^ix^j}$, $p^a_{y^iy^j}$, $i\neq 
j$) 
in branes-$a$ and $a'$ ($a=1,..,6$) are equal 
in magnitude and opposite in sign. Their contributions to the 
7-brane tadpoles are also equal and opposite, and such
tadpoles cancel between pairs of brane-$a$ and
brane-$a'$ ($a=1,..,6$). Thus, one is left with 
non-zero contributions to the 7-brane tadpoles from branes 1-6
(and $1'$-$6'$)
only along the diagonal directions $[x^1y^1]$, $[x^2y^2]$ and 
$[x^3y^3]$, which then cancel with the opposite contributions from 
branes 7-9.
To show the tadpole cancellation
explicitly and to find out the resulting 
stabilized values of the complex structure and K\"ahler class moduli, 
we
choose the ($m, n$) quantum numbers along various branes  as given 
in Appendix B. These values give the same magnetic fluxes for the 
branes 1-6 as in 
eqs.~ (\ref{soln.-flux1})-(\ref{soln.-flux2}). 
On the other hand, the magnetic fluxes for branes 
$1^\prime$-$3^\prime$ are given by: 
\be
\begin{pmatrix}p'^1_{x^1y^2}\cr p'^1_{x^2y^1}\cr p'^1_{x^1y^1}\cr 
p^1_{x^3y^3}\end{pmatrix} =
\begin{pmatrix}p'^2_{x^2y^3}\cr p'^2_{x^3y^2}\cr p'^2_{x^2y^2}\cr 
p'^2_{x^1y^1}\end{pmatrix} =
\begin{pmatrix}p'^3_{x^3y^1}\cr p'^3_{x^1y^3}\cr p'^3_{x^3y^3}\cr 
p'^3_{x^2y^2}\end{pmatrix} = 
\begin{pmatrix} 1\cr  1\cr -2\cr -1\end{pmatrix}.
\label{soln.-flux1'}
\ee
Similarly, for branes $4^\prime$-$6^\prime$ the non-zero fluxes read:
\be
\begin{pmatrix}p'^4_{x^1x^2}\cr p'^4_{y^1y^2}\cr p'^4_{x^1y^1}\cr 
p'^4_{x^3y^3}\end{pmatrix} =
\begin{pmatrix}p'^5_{x^2x^3}\cr p'^5_{y^2y^3}\cr p'^5_{x^2y^2}\cr 
p'^5_{x^1y^1}\end{pmatrix} =
\begin{pmatrix}p'^6_{x^3x^1}\cr p'^6_{y^3y^1}\cr p'^6_{x^3y^3}\cr 
p'^6_{x^2y^2}\end{pmatrix} = 
\begin{pmatrix}  1\cr  1\cr -2\cr -1\end{pmatrix}.
\label{soln.-flux2'}
\ee

We can now discuss the stabilization of the complex structure and 
K\"ahler class 
moduli for model-B, specified by branes 1-6, $1^\prime$-$6^\prime$ 
and 7-9.
These 
branes alone stabilize the moduli in the present case, as well,  
to the same values:
\be
    \tau^{ij} = 0,\;\;(i\neq j),\;\;\;
    \tau^{11} = \tau^{22} = \tau^{33} = i. 
\label{complex-structure-B}
\ee
\be
  J_{i\bar{j}} = 0,\;\;(i\neq j),\;\;\;
J_{x^1y^1} = J_{x^2y^2} = J_{x^3y^3} = (2\pi)^{2}\alpha'.
\label{kahler-B}
\ee

However, one now has the additional branes $1^\prime$-$6^\prime$  and 
we must therefore 
make sure that their presence maintains the moduli 
stabilization values  (\ref{complex-structure-B}), (\ref{kahler-B}). 
To see that this is indeed the case, we first notice that  the 
values of the complex structure given from eqs.~ 
(\ref{M20_condition}) and (\ref{complex-structure-B}) remain 
invariant if one
changes the sign of all the magnetic flux components of the type, 
$p^a_{x^iy^j}$, $p^a_{x^ix^j}$, $p^a_{y^iy^j}$, $i\neq j$, 
while keeping the diagonal fluxes 
$p^a_{x^iy^i}$ unchanged. Since this
is precisely the change induced in  branes $1^\prime$-$6^\prime$, we 
conclude 
that  model-B gives still  the same solution 
for the complex structure moduli as in
eq.~(\ref{complex-structure-B}).
Next, we note that the supersymmetry conditions, written for branes
1-9 in eqs.~(\ref{kahler-1})-(\ref{kahler-9}), are respected by
the branes $1^\prime$-$6^\prime$ as well. More precisely, 
the r.h.s. of eqs.~(\ref{kahler-1})-(\ref{kahler-6}), as well as
eqs.~(\ref{susy-flux1}), written for branes $1^\prime$-$6^\prime$,
are identical with those for  branes 1-6. 
Similarly,  eqs.~(\ref{positivity-1}) and (\ref{positivity-2}),
imposing the positivity condition (\ref{Jcond}), 
remain also identical for  branes $1^\prime$-$6^\prime$ as for branes 
1-6. 
Finally eq.~(\ref{flux-equality}), used in determining the explicit 
value of the diagonal components of the K\"ahler moduli, also remains
intact when one replaces the branes 1-6 by  $1^\prime$-$6^\prime$.
We therefore have the solution of the K\"ahler moduli for model-B as
in eq.~(\ref{kahler-B}).

To show the cancellation of the 7-brane and 3-brane tadpoles 
in this model, we first note that the general expression for the 
7-brane 
tadpole contribution   remains the same as in 
section \ref{subsection-tadpole-A} for model-A. 
In Appendix B, we give the tadpole 
contributions from every brane and show the 7-brane tadpole 
cancellations. The 3-brane tadpole cancellation in this model 
is also  similar to the one discussed in 
section \ref{subsection-tadpole-A}. 
Each of the branes 1-6 and $1^\prime$-$6^\prime$  contributes 
$q_{3,R}^a = 1 $ to 
the 3-brane tadpole, whereas this contribution is zero for branes 
7-9. One therefore obtains the total 3-brane tadpole:
\be
\sum_{a=1}^6 q^a_{3, R} +
\sum_{a'=1}^6 q^{a'}_{3, R} +
\sum_{a=7}^9 q^a_{3, R} = 12,
\label{total-3brane-B}
\ee
if only a single brane of each stack  is used, $N_a = N_{a'} = 1$. To 
satisfy the  condition (\ref{t3}), one can for instance add four 
space filling $D3$
branes to the system, or alternatively consider 
multiple copies of some of the $D9$ branes 1-6 and 
$1^\prime$-$6^\prime$.

Moreover, in this model , it is also possible to introduce some R-R 
and NS-NS 
3-form
fluxes in order to fix the dilaton. Let us assume for instance the 
same 
configuration of
quanta as in section \ref{sec:A-3-form}. The value for the string 
coupling in
terms of the 3-form quanta is still given by 
eq.~(\ref{string-coupling2}), 
but the tadpole condition
(\ref{tad6final}) changes because of the higher contribution 
(\ref{total-3brane-B}) due to the
additional magnetized branes. The condition (\ref{tad3ff}) now  reads 
\be
- h_2^{12}f_1^{12} = 4,
\ee
and the minimal value for the string coupling is then given 
by $f_{1}^{12}= -2$ and
$h_2^{12}=2$, which corresponds to $g_s = 1$.

\section{Moduli stabilization using open and closed string 
fluxes}\label{sec:CD}

In the previous sections, we have shown in several 
examples that both complex structure and K\"ahler class moduli 
stabilization can be achieved in string theory involving 
wrapped $D9$ branes, using  magnetic fluxes that are turned on 
along the compactified directions.
In this section, we present models where some of the complex 
structure 
and K\"ahler class moduli are fixed using the 3-form fluxes that were 
introduced in sections 
\ref{sec:A-3-form} and \ref{sec:Model-B}
to stabilize the axion-dilaton field.
To  this  end, we make use of the primitivity 
condition (\ref{primitivity}) and the superpotential 
variation eqs.~(\ref{3ffcond1})-(\ref{3ffcond3}) to 
put several constraints on the geometric moduli.
The remaining ones are then fixed by the magnetic fluxes along the
branes, as in sections \ref{explicit} and \ref{sec:Model-B}.

\subsection{Model-C with $q_3 = 4$ and 3-form fluxes 
}\label{sec:model-C}

As an explicit example, we  present a model (called model-C), in 
which the  3-form fluxes involve four non-vanishing parameters of 
eq.~(\ref{dil:flux2}).
The conditions imposed by this flux on the complex structure and 
dilaton 
moduli are given in (\ref{3ffcond1})-(\ref{3ffcond3}). 
Eqs.~(\ref{3ffcond3}) give rise to five conditions on the nine 
complex structure matrix elements:
\be
\tau_{13} =\tau_{23} =\tau_{31} = \tau_{32} = 0 \quad , \quad 
f_1^{12} \tau_{21} + f_1^{21}\tau_{12} = 0.
\label{Ctau}
\ee
Condition  (\ref{3ffcond1}) is then 
trivially satisfied, while eq.~(\ref{3ffcond2}) restricts the  
flux parameters by
\be
f_1^{12}h_2^{12} = f_1^{21}h_2^{21}.
\label{Crestr}
\ee
Finally, the condition (\ref{3ffcond3}) relates the axion-dilaton 
field $\phi$ to the yet undetermined complex structure element 
$\tau_{33}$
\be
 \phi h_2^{12} \ = -\tau_{33}f_1^{21}.
\label{Cphi}
\ee

The above relations (\ref{Ctau}), (\ref{Crestr}) and  (\ref{Cphi}) 
assure 
that the
3-form flux $G_{(3)}$ is of the type $G_{(2,1)}$. If we anticipate 
the fact that the magnetic fluxes fix the remaining off diagonal 
complex 
structure component to zero, $\tau_{12}=0$, the 3-form flux reads:
\be
 -2iG_{(2,1)} = {f_1^{21} \over {\rm Im }\tau_{11}} dz^1 \wedge
 dz^{\bar{1}}\wedge dz^{3} -  {f_1^{12}\over {\rm Im}\tau_{22}} dz^2 
\wedge dz^{\bar{2}}\wedge dz^{3}.
\label{Cg21}
\ee
Let us turn on to the restriction on the K\"ahler 
form coming from the primitivity condition $G_{(2,1)} \wedge J = 0 $ 
which  is a $(3,2)$-form. 
As there exists three of them on $T^6/\mathbb{Z}_2$, this condition 
could give 
rise to a maximum of three complex conditions on the K\"ahler form. 
In our case, the choice of fluxes made in eq.~(\ref{dil:flux2}) 
restricts the K\"ahler moduli space to 
\be
J_{1\bar{3}} = J_{2\bar{3}} = 0 \quad , \quad f_1^{21} 
{J_{2\bar{2}}\over{\rm Im}
  \tau_{11}} =  f_1^{12}{ J_{1\bar{1}}\over {\rm Im}\tau_{22}}.
\label{CKcond}
\ee 
Thus, in a supersymmetric vacuum, the presence of the closed string 
fluxes 
(\ref{dil:flux2}) restricts  the metric moduli space. There 
are five complex structure and three K\"ahler class moduli which are 
fixed. They correspond to a factorized geometry of the form $T^4 
\times T^2$, where the complex structure $\tau_{11}$, $\tau_{22}$ and 
$\tau_{12}$ of the $T^4$  and  $\tau_{33}$ of the $T^2$ remains
unfixed. In the same way, the K\"ahler moduli 
$J_{1\bar{1}}$, $J_{2\bar{2}}$  and $J_{1\bar{2}}$ of the $T^4$, as 
well as the area $J_{3\bar{3}}$ of the $T^2$ are not stabilized by 
the closed 
string moduli. They correspond to four complex parameters for the 
complex structure and four real ones for the K\"ahler class. 

In order to fix the remaining moduli, we switch on internal magnetic 
fields, 
using branes 1-4 presented of 
section \ref{b&f}, with fluxes given in 
eqs.~(\ref{no.1})-(\ref{no.4}).
In this example, we also use the quantum numbers ($m, n$) for the 
branes 1-4, 
given in eqs.~(\ref{soln.-1'})- (\ref{soln.-2'}).
In addition, we use the three stacks of branes with only diagonal 
fluxes, 7-9 given in 
eqs.~(\ref{no.7})-(\ref{no.9}). 
However, the quantum numbers $(m, n)$ for these branes are now 
different from the ones in eq.~(\ref{soln.-3'}) as will be 
specified later in eqs.~(\ref{mn7-c}) and (\ref{mn9-c}).
We have already seen  in 
section \ref{model-A} that branes 1-3 
fix the ratios of the diagonal components of the complex structure,
according to (\ref{ratio}). Branes 1-4 then completely determine
all  diagonal components
$\tau^{11}$, $\tau^{22}$ and $\tau^{33}$. The presence of these 
magnetized $D9$ branes
also fixes the remaining off-diagonal component of $\tau^{ij}$
to zero. We have thus stabilized all  
complex structure moduli, using the corresponding 3-form fluxes 
(\ref{dil:flux2}) 
and branes 1-4, to the value
$\tau^{ij} = i \delta^{ij}$, as in eq.~(\ref{soln.-diagonal-tau}).

Now, to stabilize the remaining K\"ahler class moduli, we use  branes 
7-8, as well as branes 1-4. 
The corresponding supersymmetry conditions (\ref{kahler-7}), 
(\ref{kahler-8}) and  (\ref{kahler-1})-(\ref{kahler-4}) has as 
solution:
\be
J_{1\bar{1}}=J_{2\bar{2}}=J_{3\bar{3}} \,\,\, , \,\,\,
J_{1\bar{2}}=J_{2\bar{1}}=0.
\ee
Furthermore, the actual value for the  diagonal K\"ahler components 
is the same as 
in eqs.~(\ref{diagonal-J=1}).

After we have  shown the K\"ahler and complex structure moduli
stabilization for  model-C, we  can discuss the tadpole cancellation 
conditions.
In fact, brane-9 is needed only for  tadpole cancellation
and does not in any way disturb the moduli stabilization 
obtained above. The 7-brane tadpole contributions from  branes 1-4 
is then given by (using $l=-1$):
\be
 \sum_{a=1}^4 q^{a}_{[x^1y^1]} = 10,\;\;
\sum_{a=1}^4 q^{a}_{ [x^2y^2]} = 6,\;\;
\sum_{a=1}^4 q^{a}_{ [x^3y^3]}= 8 .
\label{total-7tadpoleC,1-6}
\ee
To cancel these tadpoles using branes 7-9, we  modify the values 
of the corresponding quantum numbers $(m,n)$  compared to the ones 
of eqs.~(\ref{soln.-3'}) to:
\be
\begin{pmatrix}(m^{7}_{x^1y^1}, n^{7}_{x^1y^1}) \cr
          (m^{7}_{x^2y^2}, n^{7}_{x^2y^2}) \cr
          (m^{7}_{x^3y^3}, n^{7}_{x^3y^3})\end{pmatrix} = 
\begin{pmatrix}(-1, 1) \cr (0, 6)\cr (-1, 
1)\end{pmatrix} 
\,\, , \,\,
\begin{pmatrix}(m^8_{x^2y^2}, n^8_{x^2y^2}) \cr
          (m^8_{x^3y^3}, n^8_{x^3y^3}) \cr
          (m^8_{x^1y^1}, n^8_{x^1y^1})\end{pmatrix} = 
        \begin{pmatrix}(-1, 1) \cr (0, 4)\cr (-1, 
1)\end{pmatrix}
\label{mn7-c}
\ee
\be
\begin{pmatrix}(m^9_{x^3y^3}, n^9_{x^3y^3}) \cr
          (m^9_{x^1y^1}, n^9_{x^1y^1}) \cr
          (m^9_{x^2y^2}, n^9_{x^2y^2})\end{pmatrix} = 
\begin{pmatrix}(-1, 1) \cr (0, 2)\cr (-1, 
1)\end{pmatrix}
\label{mn9-c}
\ee
We then obtain the following 7-brane tadpole contributions from these 
branes:
\be
    \sum_{b=7}^9q^{b}_{ [x^1y^1]} = -10 ,\;\;
\sum_{b=7}^9q^{b}_{ [x^2y^2]} = -6 ,\;\;
\sum_{b=7}^8 q^{b}_{ [x^3y^3]}=  -8 \, ,
\label{total-7tadpoleD,7-9}
\ee
which precisely cancel the  contributions  
(\ref{total-7tadpoleC,1-6}) from branes 1-4.

On the other hand, the total 3-brane tadpole  in this model (from 
single copies of branes 1-4, and 7-9) is equal to $q^R_3 = 4$ 
which, after adding the 3-form flux contribution should satisfy 
eq.~(\ref{t3}). Using eq.~(\ref{tad3ff}), we get:
\be
  N_{3} + N_{D3} = - f_1^{12} h_2^{12} = 12.
\label{Nf-c}
\ee
Choosing $N_{D3}=0$, corresponding to the case when no space-filling
$D3$ branes are introduced, we get for the 3-form flux: 
\be
N_{3} = -f_1^{12} h_2^{12} = 12.
\label{Nff}
\ee
Finally,  the axion-dilaton modulus is  stabilized 
by the 3-form fluxes at a  value  given  in  
eq.~(\ref{string-coupling2}).
To obtain a weak coupling string theory solution, 
we choose the maximum possible value for the dilaton modulus,
by choosing $h_2^{12} = 2$, $f_1^{12} = -6$, implying the value for 
the string coupling 
\be
g_s = {1\over 3}\, .
\label{gs-c}
\ee

\section{Conclusion}\label{sec:conc}

In this work, we presented several consistent string models 
based on $T^6/\mathbb{Z}_2$ orientifolds of type IIB theory,
having ${\cal N}=1$ supersymmetry in four dimensions and stabilized
complex structure and K\"ahler class moduli using open string 
magnetic fluxes. We have also shown that the dilaton-axion 
modulus can be stabilized by turning on closed string 3-form 
fluxes consistently with the leftover supersymmetry and the
fixed values of the geometric moduli in the presence of the
magnetic fields. By tuning the fluxes appropriately, we found
an infinite but discrete series of vacua where some radii are
fixed at arbitrarily large values, while the dilaton can be 
stabilized at arbitrarily weak values for the string coupling.

An advantage of fixing moduli using internal magnetic fields
is that the method has an exact string description and the
spectrum, as well as the effective interactions, are calculable
in terms of modified boundary conditions for the world-sheet
fields. The method has also a direct application to string model
building based on intersecting branes, while it can in principle 
be generalized to include open string moduli breaking gauge
symmetries. Finally, we have presented examples where
some of the complex structure and K\"ahler class moduli are 
stabilized by the magnetic fluxes whereas the remaining ones,
as well as the axion-dilaton, are stabilized using the 3-form fluxes.
Among interesting open problems is to study 
non-supersymmetric vacua with stabilized moduli and count
consistent solutions in this corner of the string landscape.

\section*{Acknowledgments} 

We would like to thank for useful discussions Massimo Bianchi, Ralph 
Blumenhagen, Juan Cascales,  
Hans Jockers, Elisa Trevigne and Angel Uranga. AK thanks the CERN 
Theory 
Division for warm hospitality during the course of this work. TM 
thanks the Swiss Army for kind hospitality, where part of this work 
has been done. 
This work was supported in part by the European Commission under 
the RTN contract MRTN-CT-2004-503369, and in part by the 
INTAS contract 03-51-6346.

\appendix

\section{Notations}\label{notations}

\subsection{Parametrization of $T^6$}

Consider a six-dimensional torus $T^{6}$ having six coordinates 
$u_k$, $k=1,\dots , 6$ with periodicity normalized to unity 
$x^i=x^i+1$, $y_i=y_i+1$~\cite{Moore:1998pn}. Writing the 
coordinates $u_k$ as $x^i$ , $y_i$, we choose then the 
orientation\footnote{This is  the orientation of \cite{Moore:1998pn}, 
which is different from the one of \cite{Frey:2002hf}. } 
\be
\int_{T^6} dx^1\wedge dy_1\wedge dx^2\wedge dy_2\wedge dx^3\wedge 
dy_3  = 1
\label{orientation}
\ee
and define the basis of the cohomology $H^3(T^6,\mathbb{Z})$
\bea
\alpha_0 & = &  dx^1\wedge dx^2\wedge dx^3 \nonumber \\
\alpha_{ij} & = & \frac{1}{2}\epsilon_{ilm}dx^l\wedge dx^m\wedge dy_j 
\label{H3basis} \\
\beta^{ij} & = & -\frac{1}{2}\epsilon^{jlm}dy_l\wedge dy_m\wedge dx^i 
\nonumber\\
\beta^0 & = & dy_1\wedge dy_2\wedge dy_3, \nonumber
\eea
forming a symplectic structure on $T^6$:
\be
\int_{T^6} \alpha_a \wedge \beta^b = -\delta_a^b\, \, ,
\,\,\,\, \textrm{for}\,\,  a,b =1,\cdots,h_3/2\, , 
\label{symp_st}
\ee
with $h_3 = 20$, the dimension of the cohomology 
$H^3(T^6,\mathbb{Z})$. 

We can also choose complex coordinates 
\be z^i = x^i + \tau^{ij}y_j,
\label{complex_structure}
\ee 
where $\tau^{ij}$ is a complex $3 \times 3$ matrix 
parametrizing the complex structure. In  this basis, the cohomology 
$H^3(T^6,\mathbb{Z})$ decomposes in four different cohomologies 
corresponding to the purely holomorphic parts and those with mixed 
indices:
\be
H^3(T^6) = H^{3,0}(T^6)\oplus 
H^{2,1}(T^6)\oplus H^{1,2}(T^6)\oplus 
H^{0,3}(T^6).
\ee
The purely holomorphic cohomology $H^{3,0}$ is one-dimensional and is 
formed by the holomorphic three-form $\Omega$ for which we choose the 
normalization
\be
\Omega = dz^1\wedge dz^2 \wedge dz^3. \label{omega}
\ee
In terms of the real basis (\ref{H3basis}), this can be written as 
\be
\Omega = \alpha_0 + \tau^{ij}\alpha_{ij} -{{\rm cof} 
\tau}_{ij}\beta^{ij} + \det \tau \beta_0,
\label{omegareal}
\ee
where ${{\rm cof}\tau}_{ij}$ is given by ${{\rm cof}\tau} = 
\left(\det \tau \right)
\tau^{-1,T}$.
We can then define the periods of the holomorphic 3-form to be
\be
\tau^a = \int_{A_a}\Omega\quad , \quad F_b = \int_{B^b} \Omega\, .
\ee
 Note that the period $F_b$ can be written as the derivative of a 
prepotential $F$: $F_b = \partial_{\tau^b} F$.

Similarly, the cohomology $H^2(T^6,\mathbb{Z})$ decomposes also in 
three cohomologies
\be
H^2(T^6) = H^{2,0}(T^6)\oplus 
H^{1,1}(T^6)\oplus 
H^{0,2}(T^6).
\ee
We choose the basis $e^{i\bar{j}}$ of $H^{1,1}$ to be of the form 
\be
e^{i\bar{j}} = i dz^i \wedge dz^{\bar{j}}.
\label{2fbasis}
\ee
The K\"ahler form can therefore by parametrized as 
\be
J= J_{i\bar{j}} e^{i\bar{j}}.
\ee
As the K\"ahler form is a real form, its elements satisfy the reality 
condition $J^{\dagger}_{i\bar{j}} = J_{j\bar{i}}$. Therefore $J$ 
depends only on nine real parameters.

\section{Quantum numbers $(m,n)$ in Model-B}

In this appendix we give some more details on model-B 
presented in section \ref{sec:Model-B}.
\bea
[(m^1_{x^1y^2}, n^1_{x^1y^2}), (m^1_{x^2y^1}, n^1_{x^2y^1}),
(m^1_{x^1y^1}, n^1_{x^1y^1}),(m^1_{x^3y^3}, n^1_{x^3y^3})] = \cr
\cr
[(m^2_{x^2y^3}, n^2_{x^2y^3}), (m^2_{x^3y^2}, n^2_{x^3y^2}),
(m^2_{x^2y^2}, n^2_{x^2y^2}),(m^2_{x^1y^1}, n^2_{x^1y^1})] = \cr
\cr
[(m^3_{x^3y^1}, n^3_{x^3y^1}), (m^3_{x^1y^3}, n^3_{x^1y^3}),
(m^3_{x^3y^3}, n^3_{x^3y^3}),(m^3_{x^2y^2}, n^3_{x^2y^2})] = \cr
\cr
= [(-1,1), (1, -1), (2,-1), (1,-1)] 
\label{B-mn1-3}
\eea
\bea
[(m^4_{x^1x^2}, n^4_{x^1x^2}), (m^4_{y^1y^2}, n^4_{y^1y^2}),
(m^4_{x^1y^1}, n^4_{x^1y^1}),(m^4_{x^3y^3}, n^4_{x^3y^3})] = \cr
\cr
[(m^5_{x^2x^3}, n^5_{x^2x^3}), (m^5_{y^2y^3}, n^5_{y^2y^3}),
(m^5_{x^2y^2}, n^5_{x^2y^2}),(m^5_{x^1y^1}, n^5_{x^1y^1})] = \cr
\cr
[(m^6_{x^3x^1}, n^6_{x^3x^1}), (m^6_{y^3y^1}, n^6_{y^3y^1}),
(m^6_{x^3y^3}, n^6_{x^3y^3}),(m^6_{x^2y^2}, n^6_{x^2y^2})] = \cr 
\cr
[(-1,1), (1,-1), (2,-1), (1,-1)] 
\label{B-mn4-6}
\eea
\bea
[({m'}^1_{x^1y^2}, {n'}^1_{x^1y^2}), ({m'}^1_{x^2y^1}, 
{n'}^1_{x^2y^1}),
({m'}^1_{x^1y^1}, {n'}^1_{x^1y^1}),({m'}^1_{x^3y^3}, 
{n'}^1_{x^3y^3})] = \cr
\cr
[({m'}^2_{x^2y^3}, {n'}^2_{x^2y^3}), ({m'}^2_{x^3y^2}, 
{n'}^2_{x^3y^2}),
({m'}^2_{x^2y^2}, {n'}^2_{x^2y^2}),({m'}^2_{x^1y^1}, 
{n'}^2_{x^1y^1})] = \cr
\cr
[({m'}^3_{x^3y^1}, {n'}^3_{x^3y^1}), ({m'}^3_{x^1y^3}, 
{n'}^3_{x^1y^3}),
({m'}^3_{x^3y^3}, {n'}^3_{x^3y^3}),({m'}^3_{x^2y^2}, 
{n'}^3_{x^2y^2})] = \cr
\cr
= [(1,1), (-1, -1), (2,-1), (1,-1)] 
\label{B-mn1'-3'}
\eea
\bea
[({m'}^4_{x^1x^2}, {n'}^4_{x^1x^2}), ({m'}^4_{y^1y^2}, 
{n'}^4_{y^1y^2}),
({m'}^4_{x^1y^1}, {n'}^4_{x^1y^1}),({m'}^4_{x^3y^3}, 
{n'}^4_{x^3y^3})] = \cr
\cr
[({m'}^5_{x^2x^3}, {n'}^5_{x^2x^3}), ({m'}^5_{y^2y^3}, 
{n'}^5_{y^2y^3}),
({m'}^5_{x^2y^2}, {n'}^5_{x^2y^2}),({m'}^5_{x^1y^1}, 
{n'}^5_{x^1y^1})] = \cr
\cr
[({m'}^6_{x^3x^1}, {n'}^6_{x^3x^1}), ({m'}^6_{y^3y^1}, 
{n'}^6_{y^3y^1}),
({m'}^6_{x^3y^3}, {n'}^6_{x^3y^3}),({m'}^6_{x^2y^2}, 
{n'}^6_{x^2y^2})] = \cr 
\cr
[(1,1), (-1,-1), (2,-1), (1,-1)] 
\label{B-mn4'-6'}
\eea
\bea
[(m^{7}_{x^1y^1},n^{7}_{x^1y^1}) (m^{7}_{x^2y^2},n^{7}_{x^2y^2}) 
(m^{7}_{x^3y^3},n^{7}_{x^3y^3})] &=& 
[(m^8_{x^2y^2},n^8_{x^2y^2}) (m^8_{x^3y^3},n^8_{x^3y^3}) 
(m^8_{x^1y^1},n^8_{x^1y^1})] =\cr
\cr
[(m^9_{x^3y^3},n^9_{x^3y^3}) (m^9_{x^1y^1},n^9_{x^1y^1}) 
(m^9_{x^2y^2},n^9_{x^2y^2})] &=&
[(-1,1)(0,2)(-1,1)]
\label{B-mn7-9}
\eea

\subsection{Tadpole cancellation in model-B}\label{tadpole-model-B}

The 7-brane R-R tadpole contributions, using the $(m,n)$ 
quantum numbers of eq.~(\ref{B-mn1-3}) for branes 1-3, are given as:
\be
q^{1}_{[x^1y^2]} = 1,\;\;q^{1}_{[x^2y^1]} = 1\;\; ,
q^{1}_{[x^1y^1]} = 0,\;\;q^{1}_{[x^2y^2]} = 0\;\; ,
q^{1}_{[x^3y^3]} = 1\, ,
\label{tadpole-no.1B}
\ee
\be
q^{2}_{[x^2y^3]} = 1,\;\;q^{2}_{[x^3y^2]} = 1,\;\;
q^{2}_{[x^2y^2]} = 0,\;\;q^{2}_{[x^3y^3]} = 0,\;\;
q^{2}_{[x^1y^1]} = 1\, ,
\label{tadpole-no.2B}
\ee
\be
q^{3}_{[x^3y^1]} = 1,\;\;q^{3}_{ [x^1y^3]} = 1,\;\;
q^{3}_{[x^3y^3]} = 0,\;\;q^{3}_{ [x^1y^1]} = 0,\;\;
q^{3}_{[x^2y^2]} = 1\, .
\label{tadpole-no.3B}
\ee
Similarly, for branes $1^\prime$-$3^\prime$ the expressions are:
\be
q^{1'}_{ [x^1y^2]} = -1,\;\;q^{1'}_{[x^2y^1]} = -1\;\; ,
q^{1'}_{ [x^1y^1]} = 0,\;\;q^{1'}_{ [x^2y^2]} = 0\;\; ,
q^{1'}_{ ([x^3y^3]} = 1\, ,
\label{tadpole-no.1'B}
\ee
\be
q^{2'}_{ ([x^2y^3]} = -1,\;\;q^{2'}_{ [x^3y^2]} = -1\;\; ,
q^{2'}_{ ([x^2y^2]} = 0,\;\;q^{2'}_{ [x^3y^3]} = 0\;\; ,
q^{2'}_{ ([x^1y^1]} = 1\, ,
\label{tadpole-no.2'B}
\ee
\be
q^{3'}_{ [x^3y^1]} = -1,\;\;q^{3}_{ [x^1y^3]} = -1\;\; ,
q^{3'}_{ [x^3y^3]} = 0,\;\;q^{3}_{ [x^1y^1]} = 0\;\; ,
q^{3'}_{ [x^2y^2]} = 1\, .
\label{tadpole-no.3'B}
\ee
In a similar way, one can write the contributions for branes 4-6:
\be
q^{4}_{ [x^1x^2]} = 1,\;\;q^{4}_{[y^1y^2]} = 1\;\; ,
q^{4}_{ [x^1y^1]} = 0,\;\;q^{4}_{[x^2y^2]} = 0\;\; ,
q^{4}_{ [x^3y^3]} = 1\, ,
\label{tadpole-no.4B}
\ee
\be
q^{5}_{ [x^2x^3]} = 1,\;\;q^{5}_{ [y^2y^3]} = 1\;\; ,
q^{5}_{ [x^2y^2]} = 0,\;\;q^{5}_{ [x^3y^3]} = 0\;\; ,
q^{5}_{ [x^1y^1]} = 1\, ,
\label{tadpole-no.5B}
\ee
\be
q^{6}_{ [x^3x^1]} = 1,\;\;q^{6}_{ [y^3y^1]} = 1\;\; ,
q^{6}_{ [x^3y^3]} = 0,\;\;q^{6}_{ [x^1y^1]} = 0\;\; ,
q^{6}_{ [x^2y^2]} = 1\, ,
\label{tadpole-no.6B}
\ee
and for branes $4^\prime$-$6^\prime$ as:
\be
q^{4'}_{[x^1x^2]} = -1,\;\;q^{4'}_{ [y^1y^2]} = -1\;\; ,
q^{4'}_{ [x^1y^1]} = 0,\;\;q^{4'}_{ [x^2y^2]} = 0\;\; ,
q^{4'}_{ [x^3y^3]} = 1\, ,
\label{tadpole-no.4'B}
\ee
\be
q^{5'}_{ [x^2x^3]} = -1,\;\;q^{5'}_{ [y^2y^3]} = -1\;\; ,
q^{5'}_{ [x^2y^2]} = 0,\;\;q^{5'}_{ [x^3y^3]} = 0\;\; ,
q^{5'}_{ [x^1y^1]} = 1\, ,
\label{tadpole-no.5'B}
\ee
\be
q^{6'}_{ [x^3x^1]} = -1,\;\;q^{6'}_{ [y^3y^1]} = -1\;\; ,
q^{6'}_{ [x^3y^3]} = 0,\;\;q^{6'}_{ [x^1y^1]} = 0\;\; ,
q^{6'}_{ [x^2y^2]} = 1\, .
\label{tadpole-no.6'B}
\ee

Adding the contributions from branes 1-6 and 
$1^\prime$-$6^\prime$, we obtain non-zero values only for tadpoles 
corresponding 
to the three diagonal directions $(x^1y^1)$, $(x^2y^2)$, $(x^3y^3)$. 
The final answer is:
\bea
\sum_{a=1}^6 N_a q_{[x_1y_1]}^a + \sum_{a'=1}^6 N_{a'} 
q_{[x_1y_1]}^{a'} = 
\sum_{a=1}^6 N_a q_{[x_2y_2]}^a + \sum_{a'=1}^6 N_{a'} 
q_{[x_2y_2]}^{a'} =   \sum_{a=1}^6 N_a q_{[x_3y_3]}^a + \sum_{a'=1}^6 
N_{a'} q_{[x_3y_3]}^{a'} = 4 \nonumber
\eea

It can then be verified that the above tadpole
contributions are cancelled by those of branes 7-9 
for the choice of quantum numbers $(m,n)$ given in 
eq.~(\ref{B-mn7-9}). Indeed, their contributions are:
\be
\sum_{a=7}^9 N_a q_{[x_1y_1]}^a  = 
\sum_{a=7}^9 N_a q_{[x_2y_2]}^a  =   \sum_{a=7}^9 N_a q_{[x_3y_3]}^a  
= -4. \nonumber
\ee

The 3-brane tadpole cancellation in this model is discussed in the 
text.



\end{document}